# Commuting to work and gender-conforming social norms: evidence from same-sex couples[♣]

Sonia Oreffice[♣]

Dario Sansone[♦]

February 21, 2022

**Abstract**

We analyze work commute time by sexual orientation of partnered or married individuals, using the American Community Survey 2008-2019. Women in same-sex couples have a longer commute to work than working women in different-sex couples, whereas the commute to work of men in same-sex couples is shorter than the one of working men in different-sex couples, also after controlling for demographic characteristics, partner's characteristics, location, fertility, and marital status. These differences are particularly stark among married couples with children: on average, about 3 minutes more one-way to work for married mothers in same-sex couples, and almost 2 minutes less for married fathers in same-sex couples, than their corresponding working parents in different-sex couples. These gaps among men and women amount to 50%, and 100%, respectively, of the gender commuting gap estimated in the literature. Within-couple gaps in commuting time are also significantly smaller in same-sex couples. We interpret these differences as evidence that it is gender-conforming social norms boosted by parenthood that lead women in different-sex couples to specialize into jobs with a shorter commute while their male partners/spouses hold jobs with a longer commute.

[♣] We thank Curtis Simon for helpful comments. All errors are our own.
[♣] University of Exeter, HCEO, and IZA. E-mail: s.oreffice@exeter.ac.uk
[♦] University of Exeter and IZA. E-mail d.sansone@exeter.ac.uk

# 1. Introduction

The gender commuting gap along with the broader gender inequality in labor market outcomes have been the focus of recent literature and of social and political concerns about persistent inequities on the basis of gender. Gender roles affecting couples' specialization in the labor market and in the household, and thus job characteristics such as commuting, flexibility, and long hours, feed into the still sizable inequality in labor market outcomes of men and women (Bang 2021; Bertrand 2020; Bertrand et al. 2021; Goldin 2021; Rodríguez-Planas and Tanaka 2022). Specifically, this inequality is estimated to be larger with parenthood, when gender-conforming norms and the extent of job demands become more salient (Angelov, Johansson, and Lindahl 2016; Bertrand, Goldin, and Katz 2010; Borghorst, Mulalic, and Ommeren 2021; Kleven, Landais, and Søgaard 2021).

Indeed, in recent research analyzing commuting decisions, (Barbanchon, Rathelot, and Roulet 2021) show that women choose jobs with less commute and that they are willing to trade-off shorter commutes with wages, while (Lundborg, Plug, and Rasmussen 2017) estimate that women move to lower-paying jobs closer to home after having children. Relatedly, (Black, Kolesnikova, and Taylor 2014) find that in US metropolitan areas with longer commutes, married women exhibit a lower labor force participation, and (Farré, Jofre-Monseny, and Torrecillas 2020) that higher average commuting time across US cities is associated with lower labor force participation of married women, especially of those with children or exposed to more gendered social norms. (Marcén and Morales 2021) report instead that a culture with more gender equality decreases the gender commuting gap of parents in the US.

Interestingly, in the literature analyzing the relationship between fertility, labor market outcomes, and intra-household specialization, researchers have estimated that the child penalty and the degree of specialization among same-sex couples is much lower than for different-sex couples. This suggests that gender-conforming social norms may be less relevant among same-sex couples. Indeed, (Andresen and Nix 2022) show that the child penalty is much lower for women in same-sex couples and indicate gender norms and preferences as main mechanisms, while occupational flexibility alone cannot explain the child income penalty experienced by heterosexual women in Norway. (Evertsson, Moberg, and Vleuten 2021) analyze the child penalty in terms of income trajectories and reveal that in Scandinavian countries the social construction of gender and identity theory is much stronger in different-sex than in female same-sex couples. Moreover, (Hofmarcher and Plug 2022) report that in the US, same-sex couples specialize less than traditional different-sex couples, while (Giddings et al. 2014) document that over time the specialization gap between same-sex and different-sex couples has declined in the US, and (Oreffice 2011) that same-sex and different-sex couples are similarly engaged and functioning in terms of intra-household bargaining in their labor supply decisions. Nevertheless, (Jepsen and Jepsen 2022) estimate that women in same-sex couples work more hours per week, and men fewer hours, than married women and men in different-sex couples as recently as 2019 in the US.

We present novel evidence on commuting by sexual orientation, comparing the work commute behavior of same-sex couples and different-sex couples, using data from the American Community Survey (ACS) 2008-2019. This helps us disentangle possible mechanisms behind the well-known gender commuting gap and labor inequalities on the basis of gender: are these gender differences mainly due to biological differences between men and women, or to gender-conforming social norms that induce men and women to take up different roles in the household and in the labor market, especially when parenthood arises (Bertrand, 2020)? Men and women across types of couples share the same biological sex and couple status, however same-sex couples may not be subject to the same degree of gender-conforming social norms that push women in different-sex household to settle for less demanding (rewarding) and closer-to-home jobs, while their men settle for farther, more demanding (rewarding) jobs.

Although there is a lack of (large) datasets containing information on travel time to work, labor market outcomes and sexual orientation, same-sex couples can be identified in the ACS by matching household heads with their same-sex spouses or unmarried partners. We exploit the availability of the variable "Travel time to work", reporting the total amount of time, in minutes, that it usually took the respondent to get from home to work, in the week preceding the survey interview. This information is available for the respondent but also for their unmarried partner or spouse, if present and working. We build the most recent and largest sample with detailed demographic, labor, and travel time to work information on respondents and their partners or spouses in same-sex couples, along with standard samples of respondents and their partners or spouses in different-sex couples, focusing on employed adult individuals 18-64 years old.

We find that working women in same-sex couples commute longer to work than working women in different-sex couples, whereas the commute to work of working men in same-sex couples is shorter than of working men in different-sex couples. These differences persist after controlling for demographic characteristics, partner's characteristics, location, fertility, and marital status, and are particularly stark among married couples with children: on average, almost 3 minutes more one-way to work for married mothers in same-sex couples, and almost 2 minutes less for married fathers in same-sex couples (this gap is also larger than 2 minutes for unmarried fathers). Within-couple commuting gaps are also smaller in same-sex couples. Our estimates are robust to focusing on secondary-earners only, main earners only, household heads, partners/spouses only, individuals forty years old or older, controlling for occupation or industry, family income, or to excluding those working from home.

These disparities by sexual orientation are sizable and represent an 11% increase with respect to the average commuting time of women in different-sex couples (12% among parents), and a 5% decrease with respect to the average commuting time of men in different-sex couples. The figure for women actually corresponds to the whole gender commuting gap estimated in the literature to be about 10%, while for men it is half as large but still substantial (Crane 2007). Reassuringly, when we additionally focus on the hours worked of men and women across types of couples we find the same patterns as with commuting to work: working women in same-sex couples work longer hours per week than working women in different sex couples, whereas working men in same-sex couples work fewer hours per week than working men in different-sex couples; the largest labor supply gaps exist among women and men in married couples with children, and the disparity by couple type is always of opposite sign by sex, regardless of marital or parental status.

Our findings are consistent with gender-conforming social norms internalized into preferences nudging women in different-sex couples to earn less and commute less, especially in the presence of children. Couples younger than 40 are more similar across sexual orientations as different-sex couples may have become more egalitarian in recent years and suffer less from conforming to stereotypes (Giddings et al. 2014). We show that gender stereotypes do shape women's choices in different-sex couples, but are not necessarily present among individuals in same-sex couples. Indeed, we estimate a commuting gap among men that is opposite to the gap among women, and larger in absolute value for couples with children. Men in same-sex couples have a shorter commute, but the disparity is smaller than among women. The fact that the larger estimated difference by sexual orientation is among women supports the social norms explanation that the pressure of work-family balance is primarily on women, who settle into less demanding (rewarding) jobs associated with a shorter commute to accommodate family duties (Crane 2007). These gender-conforming norms that are reinforced by parenthood are clearly present among working couples (Bertrand 2020; Kleven, Landais, and Søgaard 2021), rather than among traditional couples only, as instead suggested by (Hofmarcher and Plug 2022).

These differences in commuting to work by sexual orientation are not consistent with biological effects or with general constraints imposed by parenthood, because we consider individuals in same-sex couples who are parents as well. Our findings are in line with the rejection in the literature of the biological channel as an explanation for the child penalty suffered by women in the labor market (Andresen and Nix 2022; Kleven, Landais, and Søgaard 2021). We also analyze couples who live in the city center separately from those who do not, ruling out that these commuting gaps arise because men or women in same-sex couples live in high-amenity places in city centers instead of the suburbs where different-sex couples with young children usually live in the US. Our commute gaps hold also for those living in the city center, where women in same-sex couples still exhibit longer commutes to work than working women in different-sex couples, and men in same-sex couples shorter ones.

(Smart, Brown, and Taylor 2017) implement a strategy similar to ours in the American Time Use surveys (ATUS) from 2003 to 2012 to measure travel time across types of households. They find that household-related travel time of same-sex couples lies in between men's and women's travel time in different-sex couples. However, their sample has only 133 men and 168 women in same-sex couples, and considers years in which same-sex couples could not be classified as married in the data. Furthermore, the ATUS only has information about one of the members of the couple, rather than about both as in the ACS data we use, and controls used in their analysis are not provided. (Rapino and Cooke 2011), on the other hand, use IPUMS Census 2000 data on commuting time, and find that cohabiting same-sex couples of men and women commute the same, and that women in cohabiting same-sex couples commute the same as married men in different-sex couples: however, they cannot distinguish marital status from sexual orientation and they use only one year of data from 20 years ago from a source rich of different-sex couples misclassified as same-sex couples (Black et al. 2007). (Oreffice and Sansone 2022) examine differences in mode of transportation to work between same-sex and different-sex couples in the ACS: working individuals in same-sex couples are significantly less likely to drive to work than working men and women in different-sex couples, but more likely to use public transport, walk, or bike to work.

## 2. Data Description and Methodology

Our dataset is the version of the ACS publicly available through IPUMS-USA (Ruggles et al. 2021). The ACS is a nationally-representative repeated cross-section that has been conducted every year since 2000 in the US. It contains demographic, economic, social, work, and housing information. Since 2005, it has included a 1% random sample of the US population. Although the ACS does not contain direct questions on sexual orientation, it is possible to identify unmarried same-sex couples living together, and legally married ones since 2012. Indeed, household members can be classified as "unmarried partners" when recording their relationships to the household head, because roommates and unmarried partners are treated as two separated categories. Since 2012, same-sex couples have been allowed to report their actual marital status (between 2000 and 2012, same-sex married spouses were imputed as unmarried partners).

Unmarried "heads" and "unmarried partners", married "heads" and "spouses" were extracted from the ACS data using the variable "relationship to household head". The household head is defined as the person who owns or rents the house, apartment, or mobile home (if there is no such person, the first person listed can be any adult living in the household). Using the variable "sex",[1] couples with the head and the unmarried partner (or the spouse) sharing the same sex were then classified as same-sex couples, and those of different sex as different-sex couples.

We use data until 2019, discarding the 2020 wave because the COVID-19 pandemic disrupted the 2020 ACS data collection and affected data quality in 2020 (Daily et al. 2021). We start from 2008 because the US Census Bureau implemented several changes between 2007 and 2008 to reduce the number of different-sex couples misclassifies as same-sex couples (due to reporting errors in the sex question), which resulted in more reliable estimates and identification of same-sex couples. We drop observations with imputed sex or relation to the household head from our sample to further reduce such measurement errors, following common practice in this literature (Black,

---

[1] The ACS survey does not allow us to distinguish between sex and gender.

Sanders, and Taylor 2007; Oreffice 2011). Notwithstanding these issues, the US Census and the ACS remain the largest and most reliable data on same-sex couples (Sansone 2019). These IPUMS-USA data sources have also been commonly used for decades in urban planning and transportation studies on the gender commuting gaps (MacDonald 1999).

We focus on employed adults aged 18 to 64 who worked the week before the survey interview. All variables used in our empirical analysis are described in Section A of the Online Appendix and in Table B3, while Tables B1-B2 report sample sizes by year, sex, couple type, and marital status. As previewed in the Introduction, our main variable of interest is "Travel time to work", reporting the total amount of time, in minutes, that it usually took the respondent to get from home to work, in the week preceding the survey interview for all individuals who worked during that week. This information is available for the respondent but also for their unmarried partner or spouse, if present and working in the week preceding the interview. The commuting time of those working from home is set to zero.

The following equation is estimated for each individual *i* living in state *s* at time *t*:

$$y_{ist} = \alpha + \beta SSC_{ist} + \delta_s + \mu_t + \gamma X_{ist} + \varepsilon_{ist}$$

The main empirical specification estimates an OLS regression model where the dependent variable $y_{ist}$ is the time in minutes of a one-way commute to work for individual i living in state s at time t. Most of the empirical analysis examines whether and how a binary indicator for being in a same-sex couple ($SSC_{ist}$) is associated to this commuting time to work. The other main regressors are state and year fixed effects ($\delta_s$ and $\mu_t$), and the individual-level controls ($X_{ist}$): the respondent's age, race, ethnicity, and education, their partner/spouse's characteristics, the couple's marital status and the number of own (total and younger than 5) children living in the household. Standard errors clustered at the household level are used throughout, as well as individual weights. In our sensitivity analysis, we add a set of dummy variables for occupation or industry, control for family income, urbanicity, or homeownership. We also restrict our sample to secondary-earners only, main earners only, household heads, partners/spouses only, by race and ethnicity, age groups, to those who do not work from home, or to dual-earner couples.

We then compute the variable commuting gap as an additional outcome of interest, defined as the absolute difference of the commuting time in minutes within a couple, to further measure intra-household specialization and whether same-sex couples are more egalitarian or not.

In addition, we run standard labor supply regressions for men and women separately, where hours worked are defined as the number of hours that an individual usually works per week in the 12 months preceding the ACS interview. We use the same regression specifications as in our commuting analysis, except for adding the control for hourly wage. As sensitivity analysis, we include commuting time as a regressor. We can thus investigate whether the sexual orientation gaps that men and women exhibit in commuting to work are also present in the hours worked by men and women living in a same-sex and a different-sex couple.

Finally, we run standard wage regressions for men and women separately, where wage is defined as the hourly wage, and include a dummy for type of couple; we then add commuting time to our usual set of regressions, to explore whether the sexual orientation wage gaps estimated in the literature may be at all related to the couples' commuting gaps.

## 3. Results

### 3.1 Descriptive Statistics

Table 1 and Figure 1 reports the summary statistics of our sample by type of couple and sex in terms of commuting patterns. Working men systematically have longer work commutes than working women do, on average a difference of 4 minutes on a one-way journey from home to the workplace. When we break down this gap to distinguish between individuals in same-sex couples and individuals in different-sex couples, we find that commuting disparities by gender are more nuanced: on average, working women in same-sex couples have a 2.5-minute longer commute than working women in different-sex couples (when counting those working from home as zeros), whereas working men in same-sex couples have a 1.5-minute shorter commute than working men in different-sex couples (Table 1 row 1 and Figure 1 panel A). Given that 4 minutes represent the average gender commuting gap in the sample, these gaps by sexual orientation are sizable.

These differences by gender and couple type also emerge from the probability density functions plotted in Figure 2: women in different-sex couples are more concentrated at lower commuting times, and the gap between women in same-sex and different-sex couples is more visually established than the one between men in same-sex and different-sex couples. These conclusions remain qualitatively similar when excluding individuals working from home (Table 1 row 2 and Figure 1 panel B). These gaps in commuting time by couple type also seem not to vary substantially across years in the 2008-2019 period considered in this study (Figure B1).

Women in same-sex couples are also more likely to have long commutes than women in different-sex couples: they are almost 5 percent points more likely to have a one-way 15-minute commute to work, a similar higher probability of having a 30-minute commute, and a 2 percentage points higher probability of having a 60-minute commute. Men in same-sex couples have instead a 1-2 percentage points lower likelihood of having such long commutes (Table 1 rows 3-5 and Figure 1 panels C-D). Furthermore, commuting differences by sexual orientation are larger among parents than in households with no children (Table 1 rows 6-7). However, it is worth noting that men in same-sex couples commute longer than women in same-sex couples: the gender commuting gap holds also among individuals in same-sex couples (Table 1).

Measuring the within-couple commuting gap reveals that the difference in same-sex couples' commute times to work are more similar than in different-sex couples, and particularly for women, the disparity within couples is almost 2-minute shorter when the woman lives in a same-sex couple (Table 1 row 8 and Figure 3 Panel A). The probability density functions depicted in Figure B2 clearly show that female same-sex couples are more egalitarian and are concentrated at lower levels of within-couple commute gaps. When looking instead at the total commuting time of the two members of the couple, the gender gap in commuting time indicates that couples with two men have the highest overall commuting time, followed by couples with two women, and then by different-sex couples (Table 1 row 9 and Figure 3 Panel B).

Finally, considering the other variables in our ACS sample augmented with individuals not working and thus with missing values for commuting time, men and women in same-sex couples are on average younger, more educated, more likely to be white, less likely to have children or be married, and – at least for women – more likely to be employed, than men and women in different-sex couples (Table B3). This is in line with what previous literature on sexual orientation has documented in the US (Badgett, Carpenter, and Sansone 2021; Oreffice 2011). Among those working, the number of weekly hours worked is higher for women in same-sex couples than those in different-sex couples (41 hours/week versus 38 hours/week on average), while it is lower for men in same-sex couples than men in different-sex couples (42 versus 44 hours/week).

**3.2 Regression Analysis of Commuting: Main Results**

Table 2 reports the main regression results of commuting time to work in minutes on a binary indicator for being in a same-sex couple, separately for working women (Panel A) and working men (Panel B). Starting from the basic correlation in Column 1, controls are incrementally added, from state and year fixed effects (Column 2), to the respondent's age, race, ethnicity, and education (Column 3), their partner/spouse's characteristics (Column 4), their marital status and the number of own – total and younger than 5 – children living in the household (Column 5).

Being in a same-sex couple is associated with opposite commuting patterns for men and women: women in same-sex couples commute longer to work than women in different-sex couples, 2.5 minutes more one way on average, whereas men in gay couples have a shorter commute to work than men in heterosexual couples, 1.4 minutes less on average (Column 1). The fact that men exhibit an opposite commuting behavior to women by couple type is consistent with the idea that same-sex couples may be more egalitarian (Badgett, Carpenter, and Sansone 2021) and especially with commuting decisions being shaped by gender-conforming social norms and different household roles on the basis of sex. These gendered social expectations are much weaker in same-sex couples, and women especially benefit from this, implementing commuting work patterns more similar to men.

The mean commuting time to work is 23.2 minutes for women and 27.7 minutes for men in our overall sample: the 2.5 minute increase in commute time to work among working women in same-sex couples (Column 1) represents an 11% increase with respect to women in different-sex couples, and the 1.4 minute decrease (Column 1) among men represents a 5% decrease with respect to men in different-sex couples. Even more striking are the comparisons of these differences by sexual orientation to the 10% gender commuting gap estimated in the literature: among working women, the 11% disparity is as sizable as the whole gender commuting gap, while among working men the 5% gap is half as large. These estimated coefficients indicate a relevant commuting pattern by sexual orientation for men and especially for women.

According to the US Census Bureau (U.S. Census 2021), the average one-way commute time is at all-time high in 2019, and from 2006 to 2019 it increased by about 2.6 minutes. According to (Crane 2007), the gender gap in commute time was 2.4 minutes in 2005 in the US. These policy-relevant figures correspond to our main commuting disparity by sexual orientation among women. We also note that the measure of commuting time recorded in the ACS data is one-way commuting time: thus, the average daily difference in total commute to and from work among women would be 5 minutes, which in turn is 25 minutes per week, on average.

All these disparities by sexual orientation that we have uncovered are significant at the 1% level, and robust to controlling for demographic characteristics, partner's characteristics, fertility, and marital status, although their magnitude decreases from columns 1 to 5. In this last column, working women in same-sex couples exhibit a difference of 1.8 minutes more in their one-way commute to work than working women in different-sex couples, while for working men in same-sex couples the difference is 1 minute less than working men in different-sex couples.

Table 3 illustrates these commuting differences by sexual orientation and sex, but separately by marriage and parenthood. The dummy variable for same-sex couples is always statistically significant at the 1% level. These estimates reveal that the largest gaps exist among women in married couples with children: among this sub-group, women in same-sex couples commute almost 3 minutes longer than wives in different-sex couples, while for men the difference is 1.7 minutes less (Column 1). The disparity by couple type is always of opposite sign by sex, regardless of marital or parental status. The pattern of results is consistent with married couples specializing more than other household types (Jepsen and Jepsen 2015).

Furthermore, when comparing couples with and without children in the household (columns 1 and 3 to 2 and 4), we find that the commuting difference associated to same-sex couples is always larger in couples with children than in those without children. Commuting to work decisions also reflect couples' decisions related to fertility: couples (married or cohabiting) with no children exhibit the smallest commuting disparity by sexual orientation, supporting our interpretation that the prevalent gender commuting gap reflects gender-conforming social norms reinforced by parenthood (Borghorst, Mulalic, and Ommeren 2021; Farré, Jofre-Monseny, and Torrecillas 2020). Indeed, same-sex couples may be more egalitarian and less subject to strong division of labor and work-family balance constraints than different-sex couples are (Andresen and Nix 2022;

Evertsson, Moberg, and Vleuten 2021). Our findings do not support a biological difference explanation among types of couples. Similarly, we find the smallest difference in commuting time between same-sex and different-sex couples among unmarried couples without children: among cohabiting couples without children, women in different-sex couples may feel less pressure to adhere to gender social norms.

**3.3 Regression Analysis of Commuting: Heterogeneity**

Table 4 presents the same regression analysis as in Table 2, but on sub-samples of household heads, partners/spouses, main earners, or secondary earners in the couple, as well as by metropolitan status. The larger same-sex couple differences in commuting time by household role are associated with secondary earners among women (column 4), and with partners/spouses among men (column 2). Given that many women in different-sex couples are secondary earners in their household, the larger commuting gap with women in same-sex couples is consistent with our gender-conforming norms interpretation. Similarly, the small gap between men in same-sex and different-sex couples found when focusing on secondary earners may reflect the fact that different-sex couples in which the man is not the primary earner could be less likely to follow gender-conforming social norms.

The last two columns of Table 4 report the difference by couple type in commuting time to work among couples living in city centers (column 5) or those who don't (column 6). The specification estimated in column 5 is the same regression specification of column 5 of Table 2 but run on the restricted sample of city dwellers, who represent about 10% of the main sample. Among men in the city, the difference by couple type is very similar to the full sample; if anything, men in same-sex couples commute even less than men in different-sex couples when they live in city centers. This additional evidence suggest that these distinctive commuting patterns cannot be explained away by sexual minority men's preference to live in high-amenity places rather than in the suburbs (Black et al. 2002). Among women city dwellers, instead, same-sex couples still commute longer to work than women in different-sex couples, but the estimated difference amounts to 1 minute. It is possible that women in different-sex couples that choose to live downtown exhibit work and commuting patterns that are less gendered or less dictated by social norms and household specialization by sex (Costa and Kahn 2000; Simon 2019).

In column 6, the regression specification is the same as in column 5 of Tables 2 and 4, but run on the subsample of couples who do not live in city centers. The gap by sexual orientation among women is larger than in the full sample, and much larger than for city dwellers. Gender-conforming norms may be more salient for heterosexual women who embrace life in the suburbs, which is a traditional family living location choice among American households. On the contrary, among those couples living outside of the city center there is no difference between men in same-sex couples and men in different-sex couples: it may be the case that men in same-sex couples who decide to live in the suburbs exhibit a more traditional way of life and commute as much as men in different-sex couples.

Table 5 presents the commuting time regressions for different age groups: younger couples (aged 18 to 40), older couples (aged 41 to 64), and for our main sample but excluding individuals younger than 25. Excluding these very young couples is immaterial to our findings, whereas splitting the sample by age groups reveals much larger commuting differences among couples in their forties or older. For young couples the commuting difference among women is less than a minute, while among men it remains above one minute as in our main sample. Younger women in same-sex couples are more similar to women in different-sex couples also when controlling for number of children in the household. This is consistent with the evidence in Table 3 on commuting by parenthood, and with household decisions of older generations of different-sex couples being affected by gender-conforming norms more strongly. Parenthood does affect long-term labor market outcomes of women older than 40 in different-sex couples (Black, Kolesnikova, and Taylor 2014; Giddings et al. 2014), and this can lead to persistently shorter commutes than women in same-sex couples.

Table 6 instead splits our main sample by race and ethnicity, running the same regression analysis as in Table 2 separately for Whites, Blacks, Asian, and Hispanics. While Hispanics exhibit the same type of gaps by sexual orientation as Whites do, the commuting gap is smaller among Hispanic women, and larger among Hispanic men, than among Whites, although on average they commute 1-2 minutes longer than Whites. The commuting gaps of Blacks or Asians are not significant and are smaller, also taking into account that their commutes are much longer on average. The fact that the Hispanic commuting gap is the only significant one, and for men it is also larger, is consistent with the fact that traditionally Hispanics have stronger gender norms in place among different-sex couples.

In Table 7, we include additional controls for student status, being in the military, occupation or industry fixed effects, family income, urbanicity, homeownership, and LGBTQ+ policies. The first, fourth, fifth, sixth, and seventh columns show that the estimated differences for same-sex couples are essentially the same as those of column 5 of Table 2. However, when we control for occupation or industry fixed effects (columns 2 and 3), we observe that the significant commuting differences by sexual orientation are estimated to be smaller than in our main specification in Table 2, and for men in particular the estimated gap for same-sex couples shrinks to less than half a minute (in absolute value). Nevertheless, we note that job flexibility differences across occupations/industries or workplace locations do not explain away our findings of commuting differences by sexual orientation. Especially among women, sorting into occupations with different degrees of flexibility does not seem to be the way in which women in different-sex couples implement gender-conforming norms: the estimated coefficient associated to same-sex couples in column 2 is still 1.4 minutes and significant at the 1% level for women (Goldin 2014; Bang 2021). This is also consistent with what (Andresen and Nix 2022) find for the child penalty in Norway. Instead, men in same-sex couples seem to sort into occupations or industries that allow them to have a shorter commute to work than men in different-sex couples.

Table 8 illustrates the commuting disparities among same-sex and different-sex couples by education level of the couples. We split the sample into the four types of couples according to how high-educated (college degree or above) individuals and low-educated (less than a college degree) individuals sort into couples. Among high-educated women, those matched with high-educated partners/spouses exhibit a slightly larger commute time difference by sexual orientation (column 1), which happens to be very similar to the commuting difference among low-educated women matched with high-educated partners/spouses (column 3), but larger than the commuting differences exhibited by women in couples where both are low-educated (column 4). Instead, high-educated women matched with low-educated individuals exhibit a much smaller disparity between same-sex and different-sex women (column 2). Women matched with a high-educated man seem to respond more to work-life balance family needs than women in same-sex couples with comparable partners/spouses: both low-educated and high-educated women in different-sex couples seem to choose a shorter commute when matched with a high-educated men (column 1). It is interesting that the gender-conforming social norms do not seem salient for the uncommon couples of high-educated women and low-educated men, and are looser for low-educated women in couples where both are low-educated: the latter may specialize less or low-educated people may not be able to pick jobs and location so they are all more similar. Among men, in general the commuting differences become much smaller, except among low-educated couples. When controlling for occupation fixed effects (Table B4), these differences for men become negligible except for men in low-educated couples, with a commuting difference of only half a minute by couple type. It is also worth noting that the average commuting time for men does not change substantially across education groups, while it varies significantly for women, with high-educated women matched with low-educated partners/spouses having the longest commute, and low-educated women matched with high-educated partners/spouses having the shortest.

Our main commuting findings, as well as those from splitting the sample by education group, question (Hofmarcher and Plug 2022)'s assessment that differences between same-sex and different-sex couples are present only among traditional different-sex couples: we focus on working individuals in couples, and we still find a sizable difference by sexual orientation in commute time to work, even among couples that should be less traditional because women work in the labor market or are high-educated.

**3.4 Regression Analysis of Commuting: Robustness Checks**

Table B5 presents a battery of robustness checks confirming that same-sex couples exhibit a longer commute to work among women, and a shorter one among men. These checks include excluding students or military personnel from our sample, focusing on the 2012-2019 ACS samples, using heteroskedasticity-robust standard errors, or not using weights.

Tables B6 and B7 replicate our main regressions of Tables 2 and 3 but on the subsample of dual-earner couples: the estimated coefficients associated to being in a same-sex couple remain significant and of the same sign, of the same magnitude among women and slightly smaller among men (for men the sample size is almost halved due to the several different-sex couples where the wife does not work in the labor market). We interpret this evidence that coefficients remain very similar as supportive of social norms shaping different-sex households' work decisions: even among dual-earner couples, who should be less prone to conforming to traditional gender norms, we observe this powerful force that is instead much less present among same-sex couples.

Table B8 presents an additional robustness check: we exclude women and men who work from home from our sample, running the same set of regressions as in Table 2. Table B8 shows the same significant pattern of differences by sexual orientation in commuting time, with slightly lower magnitudes among men.

### 3.5 Regression Analysis of Within-Couple Commuting Time Gap

In Table 9 the dependent variable is the within-couple commuting time gap: the difference in minutes between the commute to work of the two partners/spouses in a couple. If members of same-sex couples have more similar work behavior and labor market outcomes, then we may expect individuals in same-sex couples to exhibit more similarities within couples also in terms of commuting time. All the specifications in Table 9 confirm this pattern: the estimated coefficient associated to being in a same-sex couple rather a different sex couple is always negative for this gap, among men and women: the commuting time within female same-sex couples is more similar by almost two minutes, whereas for male same-sex couples is more similar by less than half a minute in the richer specification of column 5. These stronger similarities within same-sex couples are consistent with gender-conforming norms affecting different-sex couples' work and household decisions, and their stronger role in shaping especially women's work choices (Bertrand 2020).

Table B9 reports the regressions separately by marital status and fertility for these within-couple differences in commuting time. It is interesting to see that the within-couple gaps are slightly more similar among parents, while among unmarried cohabiting individuals their within-couple commuting gaps are still quite different by sexual orientation.

### 3.6 Regression Analysis of Hours Worked

We now turn to compare the hours worked by same-sex couples and different-sex couples, for men and women separately. Our main analysis concerns household work choices in terms of traveling time to the workplace. If our estimated differences by sexual orientation indeed arise because of gender-conforming social norms pressuring different-sex couples, we want to verify that this mechanism also operates on the actual labor supply of men and women in the various types of couples under scrutiny, with the same set of controls and sub-samples by marital status and fertility as in our commuting analysis (plus the control for hourly wages).

Table 10 indeed reports the same pattern of results as Table 2: women in same-sex couples usually work more hours than women in different-sex couples, whereas men in same-sex couples work fewer hours per week across all specifications. The dummy variable for same-sex couples is always statistically significant at the 1% level. When we add the control for commuting time, we estimate a positive significant association between hours worked and commuting time both among men and among women, while the estimated coefficients for same-sex couples on hours worked remain significant and sizable. Being in a same-sex couple is associated to about 8% more hours worked per week for women, and 5% less for men, and these gaps go in the same direction as our estimated sizes of the commuting disparities by sexual orientation. Taken together with our commute-to-work findings, these disparities in labor supply by sexual orientation point to household work decisions being shaped by gender-conforming social norms that shift the pressure of work-family balance on heterosexual women.

Table 11 illustrates these labor supply differences by sexual orientation and sex, but separately by marriage and parenthood. These estimates reveal that the largest gaps exist among women and men in married couples with children: among these sub-groups, women in same-sex couples work almost 3.5 hours longer than wives in different-sex couples, while for men the difference is 2.3 hours less per week (Column 1). The disparity by couple type is always of opposite sign by sex, and statistically significant at the 1% level, regardless of marital or parental status. Moreover, the estimated work hour difference by sexual orientation is always larger in couples with children than in those without children, as it was the case for commuting time (Table 3). These work patterns support our interpretation that the prevalent gender commuting gap reflects gender-conforming social norms reinforced by parenthood among different-sex couples (Farré, Jofre-Monseny, and Torrecillas 2020). Indeed, same-sex couples may be more egalitarian and less subject to strong division of labor and work-family balance constraints than different-sex couples are: this is amply reflected in their choice of job characteristics such as location, in addition to hours worked.

(Jepsen and Jepsen 2022) estimate that women in same-sex couples work more hours per week, and men fewer hours, than *married* women and men in different-sex couples as recently as 2019. (Jepsen and Jepsen 2015) had previously found that in the year 2000 *married* different-sex couples specialized more than other couple types. Indeed, (Giddings et al. 2014) used the within-couple difference in hours worked in the 1990 U.S. Census and the 2000–2011 ACS data to compare same-sex to different-sex couples, and found that the former specialize less and partners are more similar than in different-sex couples (although they didn't control for wages). This labor supply evidence highlighting the role of household specialization is also in line with (Antecol and Steinberger 2013) who focused on women's labor supply in the 2000 US Census and found that the gap between women in same-sex couples and women in different-sex couples is larger when focusing on primary earners, but it is attenuated by the presence of children. However, all these papers do not distinguish between married and unmarried same-sex couples. In Norway, (Andresen and Nix 2022) estimate that among heterosexual mothers there is a sizable drop in labor supply due to children that is not present among same-sex mothers.

**3.7 Regression Analysis of Wages**

Finally, in Table 12, we measure how wages and commuting time are related, by type of couple and separately for men and women, for full-time workers. Specifically, we regress the logarithm of the hourly wage on the same-sex dummy, state and year fixed effects, to the respondent's age, race, ethnicity, and education, their partner/spouse's characteristics, their marital status and the number of own children living in the household. In columns 5-7 we also include controls for being a student, in the army, and occupation fixed effects, while adding part-time workers in columns 7-8. In all these specifications, we compare how adding commute time as regressor explains the variation in wages.

These specifications are added to test whether our estimated differences in commuting time between individuals in same-sex and different-sex couples may be related to other labor market differences by sexual orientation. In particular, a large number of studies have highlighted wage disparities by sexual orientation: men in same-sex couples (as well as single gay men) earn less and women in same-sex couples (as well as single lesbian women) earn more than their heterosexual counterparts (Klawitter 2015; Badgett, Carpenter, and Sansone 2021). Is it possible that the lesbian wage premium is partly due to the longer commutes implemented by women in same-sex couples, and that the gay wage penalty could be explained by the shorter commutes of men in same-sex couples?

Indeed, our estimated coefficient of commuting time is positively and significantly related to wages among men and among women: we find that longer commutes are associated to higher wages, consistently with the trade-off between wages and commute time recently emphasized in the literature (Barbanchon, Rathelot, and Roulet 2021). However, its magnitude is very small (0.2% of the average wage). More importantly, commuting time does not seem to explain much of the variation in wages in the working population (the inclusion of commuting time does not substantially increase the $R^2$), whereas the same-sex indicator is sizable, statistically significant, and largely unaffected by the inclusion of commuting time as additional control.

## 4. Discussion and conclusions

Our paper is related to few strands of literature: the gender differences in commuting acknowledged in economics and in the transportation and health literatures; the literature on child penalty, household specialization, and gender wage gaps more generally, the literature on sexual orientation and labor market outcomes, and the very recent literature on gender-conforming social norms and couple inequity (Bertrand 2020; Goldin 2021).

The transportation literature has long been interested in studying and measuring the sizable gender gap in commuting, showing that it is persistent over time, and discussing household roles and work/life balance as one of its main contributors (Madden 1981; Crane 2007). Until recently, less attention to commuting differences in the population was granted in the economics discipline (Barbanchon, Rathelot, and Roulet 2021).

We find that working women in same-sex couples commute longer to work than working women in different-sex couples, whereas the commute to work of working men in same-sex couples is shorter than of working men in different-sex couples. These differences are sizable, especially when compared to the gender commuting gap estimated in the literature. They are particularly stark among married couples with children, while within-couple commuting gaps are also smaller in same-sex couples. Moreover, these disparities by sexual orientation beyond the well-known gender gaps in commuting cannot be explained by women facing different job opportunities and commuting options due to their employability or labor market skills: we control for a rich set of individual characteristics such as age, educational attainment, race, ethnicity, partner's characteristics, location, fertility, and marital status, as well as for occupation or industry, family income, urbanicity, or homeownership in our sensitivity analysis. We also estimate similar gaps in labor supply by sexual orientation: not only travel time to work (job location), but the overall work time allocation decisions of different-sex couples seem to be influenced by gender-conforming social norms.

All in all, the variety of commuting behaviors and the work-hour gaps presented in this paper are consistent with social norms, reinforced by parenthood, pushing women in different-sex couples to work less and commute less than similar women in same-sex couples (the opposite holding for men). The estimated commuting gaps are larger for older individuals than for younger couples, as members of different-sex couples in younger generations may have become more egalitarian and thus more similar to same-sex couples.

Our analysis could help policy makers and especially managers and executives to tackle gender inequality in the workplace: if managers are mindful of how these gender-conforming social norms still impact women's work behavior, they may be able to implement more flexibility on the job and offer less "greedy" jobs and positions to women and mothers, with less strict office schedules (Kleven, Landais, and Søgaard 2021; Goldin 2021). Our evidence also adds to the call for policymakers to set up a strategy to weaken the gender-conforming social norms all together.

Finally, we acknowledge our study's limitations inherent to the ACS data: only LGBTQ+ individuals in same-sex partnerships or marriages can be identified (unpartnered LGBTQ+ individuals cannot), while the lack of gender identity data prevents the analysis of differences between transgender and cisgender individuals.

**Figure 1: Commuting time by sex and couple type.**

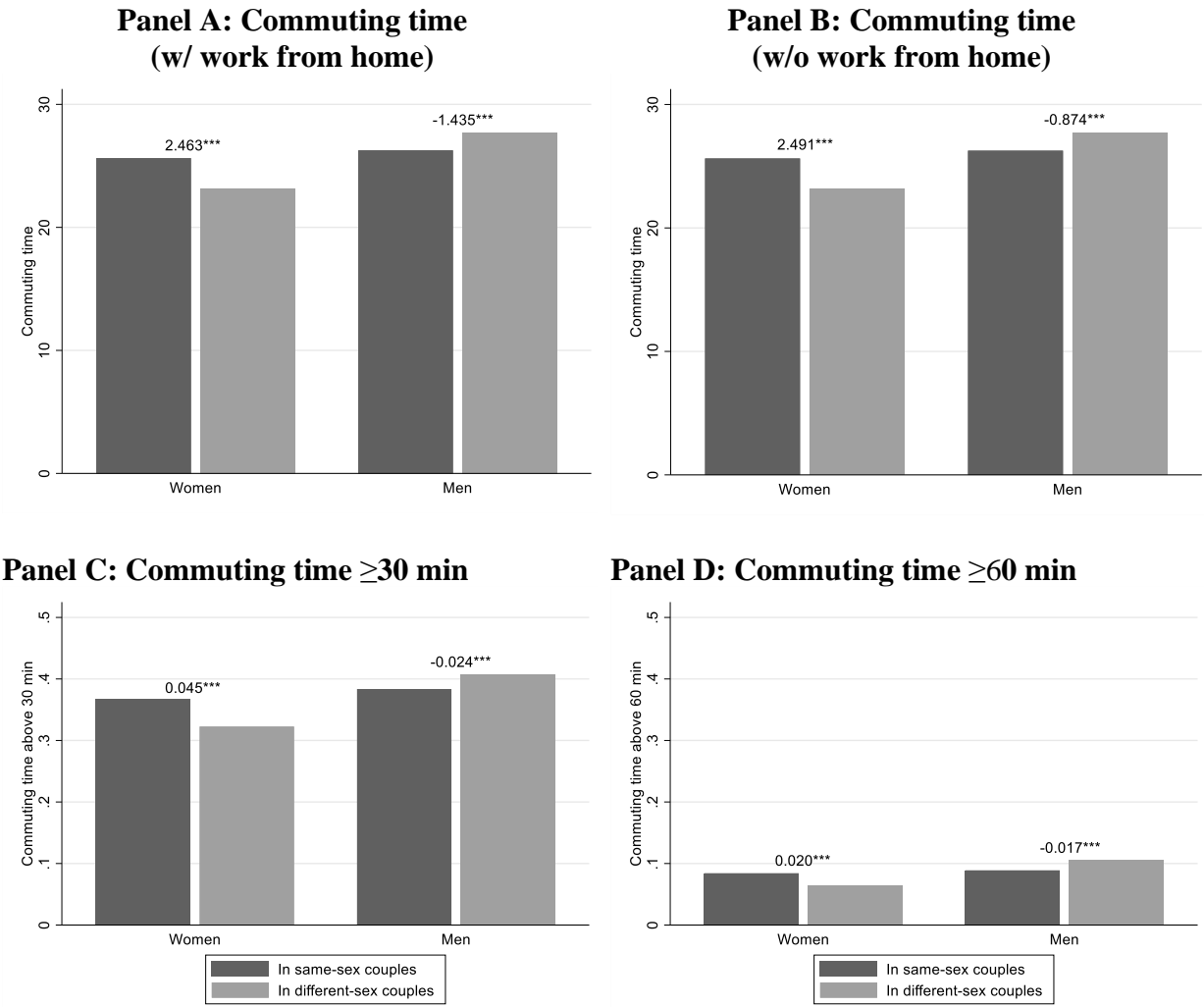

Unless otherwise specified, commuting time includes individuals working from home (commuting time imputed as zero). The number above each bar is the difference between the time for women or men in same-sex couples vs. in different-sex couples. Weighted statistics. Respondents younger than 18 or older than 64 have been excluded. Source: ACS 2008-2019. * $p < 0.10$, ** $p < 0.05$, *** $p < 0.01$.

**Figure 2: Commuting time distribution.**

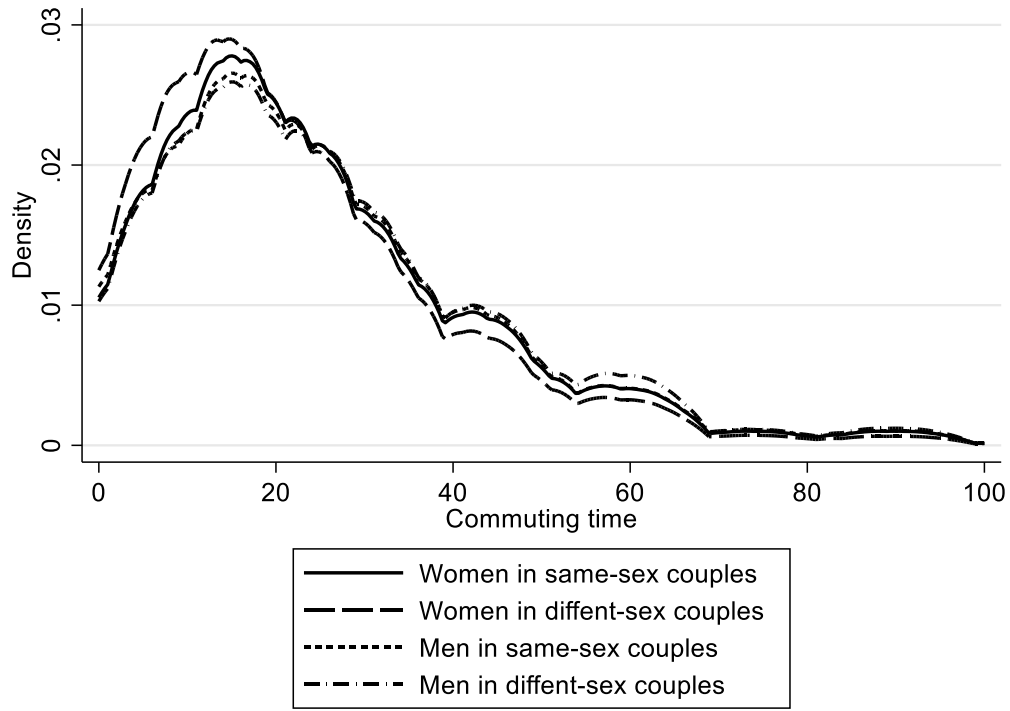

Respondents younger than 18 or older than 64 have been excluded. Commuting time censored at 100 minutes. Commuting time includes individuals working from home (commuting time imputed as zero). Bin width equal to 8 minutes. Unweighted statistics. Source: ACS 2008-2019.

**Figure 3: Couple commuting time by sex and couple type.**

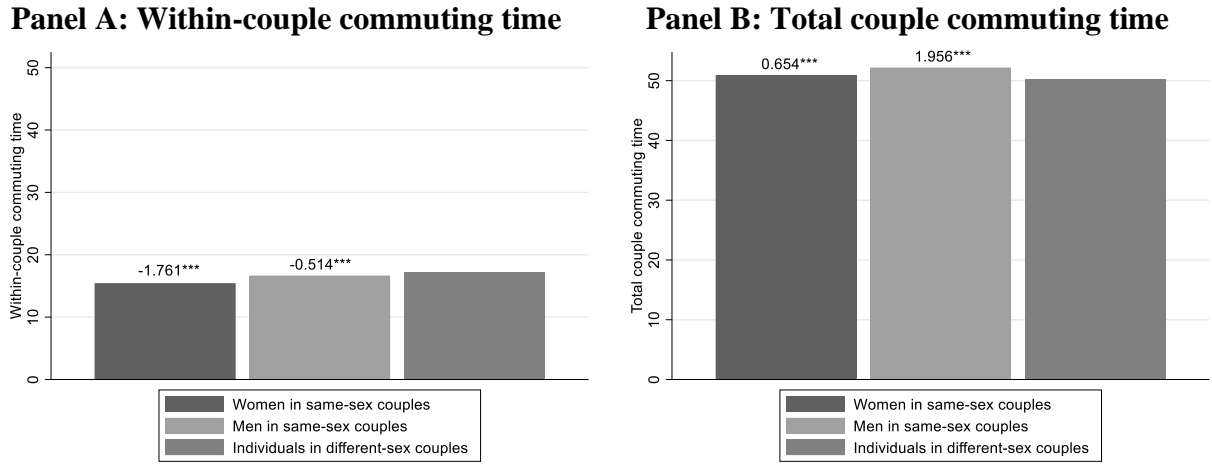

Only household heads have been included. Commuting time includes individuals working from home (commuting time imputed as zero). Couple in which at least one member was not working at the time of the survey have been excluded. The number above each bar is the difference between the time for women or men in same-sex couples vs. (women and men) in different-sex couples. Weighted statistics. Respondents younger than 18 or older than 64 have been excluded. Source: ACS 2008-2019. * $p < 0.10$, ** $p < 0.05$, *** $p < 0.01$.

**Table 1: Descriptive statistics on commuting time.**

| | | Women | | | Men | | |
|---|---|---|---|---|---|---|---|
| | | Same-sex couples | Different-sex couples | | Same-sex couples | Different-sex couples | |
| | Variable | (1) | (2) | Gap | (3) | (4) | Gap |
| 1 | Commute time | 25.626 | 23.163 | 2.463*** | 26.258 | 27.692 | -1.435*** |
| 2 | Commute time (w/o working from home) | 27.072 | 24.581 | 2.491*** | 28.168 | 29.043 | -0.874*** |
| 3 | Commute time≥15 | 0.720 | 0.675 | 0.045*** | 0.723 | 0.738 | -0.015*** |
| 4 | Commute time≥30 | 0.368 | 0.323 | 0.045*** | 0.384 | 0.407 | -0.024*** |
| 5 | Commute time≥60 | 0.084 | 0.065 | 0.020*** | 0.089 | 0.106 | -0.017*** |
| 6 | Commuting time – w/ children | 25.765 | 23.037 | 2.728*** | 26.785 | 28.283 | -1.498*** |
| 7 | Commuting time – w/o children | 25.561 | 23.344 | 2.216*** | 26.177 | 26.621 | -0.444*** |
| 8 | Within-couple commute gap | 15.409 | 17.170 | -1.761*** | 16.656 | 17.170 | -0.514*** |
| 9 | Total couple commute time | 50.877 | 50.222 | 0.654*** | 52.179 | 50.222 | 1.956*** |
| | N | 68,403 | 4,343,006 | | 66,059 | 5,144,777 | |

Unless otherwise specified, commuting time includes individuals working from home (commuting time imputed as zero). Weighted statistics. Sample size (N) refers to the total number of respondents in the relevant sub-group (i.e., those working in the week preceding the ACS interview and who reported their commuting time). See also notes in Figures 1-3. Source: ACS 2008-2019. * $p < 0.10$, ** $p < 0.05$, *** $p < 0.01$

**Table 2: Commuting time. By sex and couple type.**

|  | (1) | (2) | (3) | (4) | (5) |
|---|---|---|---|---|---|
| *Panel A: Women in SSC and DSC* | | | | | |
| In a same-sex couple | 2.463*** | 2.108*** | 2.046*** | 2.145*** | 1.761*** |
|  | (0.113) | (0.111) | (0.111) | (0.111) | (0.114) |
| Observations | 4,411,409 | 4,411,409 | 4,411,409 | 4,411,409 | 4,411,409 |
| Mean of dependent variable | 23.201 | 23.201 | 23.201 | 23.201 | 23.201 |
| $R^2$ | 0.000 | 0.020 | 0.025 | 0.026 | 0.027 |
| | | | | | |
| *Panel B: Men in SSC and DSC* | | | | | |
| In a same-sex couple | -1.435*** | -2.131*** | -1.974*** | -2.059*** | -1.021*** |
|  | (0.122) | (0.120) | (0.120) | (0.120) | (0.123) |
| Observations | 5,210,836 | 5,210,836 | 5,210,836 | 5,210,836 | 5,210,836 |
| Mean of dependent variable | 27.675 | 27.675 | 27.675 | 27.675 | 27.675 |
| $R^2$ | 0.000 | 0.019 | 0.019 | 0.019 | 0.020 |
| | | | | | |
| *Controls for:* | | | | | |
| State and year FE | | ✓ | ✓ | ✓ | ✓ |
| Demographic controls | | | ✓ | ✓ | ✓ |
| Partner/spouse controls | | | | ✓ | ✓ |
| Fertility and marital status | | | | | ✓ |

"SSC" indicates same-sex couples, "DSC" indicates different-sex couples. Commuting time includes individuals working from home (commuting time imputed as zero). Standard errors clustered at the household level in parentheses. Weighted regressions and statistics. Respondents younger than 18 or older than 64 have been excluded. *Demographic controls* include respondent's age, race, ethnicity, and education. *Partner/spouse controls* include spouse's or unmarried partner's age, race, ethnicity, and education. *Fertility* includes the number of own children (of any age or marital status) residing with the respondent, as well as the number of own children age 4 and under residing with the respondent. All variables are described in detail in Section A of the Online Appendix. Source: ACS 2008-2019. * $p < 0.10$, ** $p < 0.05$, *** $p < 0.01$.

**Table 3: Commuting time. By marital status and fertility.**

|  | Married w/ children | Married w/o children | Unmarried w/ children | Unmarried w/o children |
|---|---|---|---|---|
|  | (1) | (2) | (3) | (4) |
| *Panel A: Women in SSC and DSC* | | | | |
| In a same-sex couple | 2.765*** | 1.771*** | 1.133*** | 0.553** |
|  | (0.320) | (0.245) | (0.365) | (0.217) |
| Observations | 1,518,968 | 1,049,278 | 144,190 | 227,662 |
| Mean of dependent variable | 23.406 | 23.560 | 23.738 | 24.319 |
| $R^2$ | 0.029 | 0.026 | 0.027 | 0.032 |
| | | | | |
| *Panel B: Men in SSC and DSC* | | | | |
| In a same-sex couple | -1.662*** | -1.284*** | -2.020*** | -0.933*** |
|  | (0.498) | (0.246) | (0.767) | (0.199) |
| Observations | 1,972,381 | 1,092,622 | 166,510 | 235,897 |
| Mean of dependent variable | 28.683 | 27.156 | 27.570 | 25.945 |
| $R^2$ | 0.022 | 0.018 | 0.015 | 0.020 |
| | | | | |
| *Controls for:* | | | | |
| State and year FE | ✓ | ✓ | ✓ | ✓ |
| Demographic controls | ✓ | ✓ | ✓ | ✓ |
| Partner/spouse controls | ✓ | ✓ | ✓ | ✓ |

See also notes in Table 2. Source: ACS 2012-2019. * $p < 0.10$, ** $p < 0.05$, *** $p < 0.01$.

**Table 4: Commuting time. By position in the household and location.**

|  | Household head | Spouse or partner | Main earner | Second earner | City center | Not city center |
|---|---|---|---|---|---|---|
|  | (1) | (2) | (3) | (4) | (5) | (6) |
| *Panel A: Women in SSC and DSC* | | | | | | |
| In a same-sex couple | 1.818*** | 1.669*** | 1.050*** | 2.235*** | 0.986*** | 2.376*** |
|  | (0.146) | (0.146) | (0.126) | (0.218) | (0.251) | (0.202) |
| Observations | 1,846,540 | 2,564,869 | 1,753,489 | 2,657,920 | 452,789 | 1,991,694 |
| Mean of dependent variable | 23.318 | 23.115 | 25.084 | 21.913 | 25.702 | 24.086 |
| Adjusted $R^2$ | 0.028 | 0.026 | 0.028 | 0.026 | 0.094 | 0.030 |
| | | | | | | |
| *Panel B: Men in SSC and DSC* | | | | | | |
| In a same-sex couple | -0.910*** | -1.130*** | -0.964*** | -0.448* | -1.208*** | -0.323 |
|  | (0.155) | (0.164) | (0.137) | (0.230) | (0.220) | (0.242) |
| Observations | 3,173,588 | 2,037,248 | 4,004,896 | 1,205,940 | 547,612 | 2,336,237 |
| Mean of dependent variable | 27.422 | 28.053 | 28.044 | 26.427 | 28.066 | 29.228 |
| Adjusted $R^2$ | 0.022 | 0.018 | 0.021 | 0.021 | 0.066 | 0.023 |
| | | | | | | |
| *Controls for:* | | | | | | |
| State and year FE | ✓ | ✓ | ✓ | ✓ | ✓ | ✓ |
| Demographic controls | ✓ | ✓ | ✓ | ✓ | ✓ | ✓ |
| Partner/spouse controls | ✓ | ✓ | ✓ | ✓ | ✓ | ✓ |
| Fertility and marital status | ✓ | ✓ | ✓ | ✓ | ✓ | ✓ |

See also notes in Table 2. Column 3 includes only respondents whose individual income was greater or equal than 50% of the family income. Column 3 Panel A compares female main earners in same-sex couples to female main earners in different couples, while Column 3 Panel B compares male main earners in same-sex couples to male main earners in different couples. Columns 4 includes only respondents whose individual income was less than 50% of the family income. Column 4 Panel A compares female second earners in same-sex couples to female second earners in different couples, while Column 4 Panel B compares male second earners in same-sex couples to male second earners in different couples. Column 6 includes respondents whose metropolitan status is coded as "Not in central/principal city" or "Not in a metropolitan area", but both Columns 5 and 6 exclude respondents with undeterminable metropolitan status. Source: ACS 2008-2019.
* $p < 0.10$, ** $p < 0.05$, *** $p < 0.01$.

**Table 5: Commuting time. By age group.**

|  | 18-40 | 41-64 | 25-64 |
|---|---|---|---|
|  | (1) | (2) | (3) |
| *Panel A: Women in SSC and DSC* | | | |
| In a same-sex couple | 0.791*** | 2.183*** | 1.726*** |
|  | (0.163) | (0.154) | (0.118) |
| Observations | 1,646,346 | 2,765,063 | 4,258,026 |
| Mean of dependent variable | 23.512 | 22.985 | 23.305 |
| Adjusted $R^2$ | 0.032 | 0.026 | 0.027 |
| | | | |
| *Panel B: Men in SSC and DSC* | | | |
| In a same-sex couple | -1.407*** | -1.313*** | -1.155*** |
|  | (0.192) | (0.158) | (0.126) |
| Observations | 1,852,711 | 3,358,125 | 5,090,197 |
| Mean of dependent variable | 27.237 | 27.956 | 27.793 |
| Adjusted $R^2$ | 0.022 | 0.020 | 0.020 |
| | | | |
| *Controls for:* | | | |
| State and year FE | ✓ | ✓ | ✓ |
| Demographic controls | ✓ | ✓ | ✓ |
| Partner/spouse controls | ✓ | ✓ | ✓ |
| Fertility and marital status | ✓ | ✓ | ✓ |

See also notes in Table 2. Source: ACS 2008-2019. * $p < 0.10$, ** $p < 0.05$, *** $p < 0.01$.

**Table 6: Commuting time. By race and ethnicity.**

|  | White | Black | Asian | Hispanic |
|---|---|---|---|---|
|  | (1) | (2) | (3) | (4) |
| *Panel A: Women in SSC and DSC* | | | | |
| In a same-sex couple | 1.940*** | 0.756 | 0.029 | 1.319*** |
|  | (0.120) | (0.474) | (0.630) | (0.310) |
| Observations | 3,650,782 | 268,726 | 261,460 | 469,924 |
| Mean of dependent variable | 22.425 | 26.983 | 27.131 | 24.119 |
| Adjusted $R^2$ | 0.018 | 0.057 | 0.042 | 0.036 |
| | | | | |
| *Panel B: Men in SSC and DSC* | | | | |
| In a same-sex couple | -0.991*** | -1.044** | 0.092 | -1.805*** |
|  | (0.132) | (0.530) | (0.550) | (0.357) |
| Observations | 4,313,957 | 310,190 | 281,674 | 649,412 |
| Mean of dependent variable | 27.231 | 29.067 | 30.236 | 28.954 |
| Adjusted $R^2$ | 0.017 | 0.039 | 0.049 | 0.019 |
| | | | | |
| *Controls for:* | | | | |
| State and year FE | ✓ | ✓ | ✓ | ✓ |
| Demographic controls | ✓ | ✓ | ✓ | ✓ |
| Partner/spouse controls | ✓ | ✓ | ✓ | ✓ |
| Fertility and marital status | ✓ | ✓ | ✓ | ✓ |

See also notes in Table 2. Demographic controls in these specifications include only age and education, not race or ethnicity. Source: ACS 2008-2019. * $p < 0.10$, ** $p < 0.05$, *** $p < 0.01$.

**Table 7: Commuting time. Additional controls.**

|  | (1) | (2) | (3) | (4) | (5) | (6) | (7) |
|---|---|---|---|---|---|---|---|
| *Panel A: Women in SSC and DSC* | | | | | | | |
| In a same-sex couple | 1.754*** | 1.365*** | 1.477*** | 1.843*** | 1.742*** | 1.707*** | 1.762*** |
|  | (0.114) | (0.113) | (0.113) | (0.114) | (0.119) | (0.113) | (0.114) |
| Observations | 4,411,409 | 4,411,409 | 4,411,409 | 4,411,409 | 3,852,237 | 4,411,409 | 4,411,409 |
| Mean of dependent variable | 23.201 | 23.201 | 23.201 | 23.201 | 23.444 | 23.201 | 23.201 |
| Adjusted $R^2$ | 0.027 | 0.050 | 0.050 | 0.028 | 0.029 | 0.028 | 0.027 |
| *Panel B: Men in SSC and DSC* | | | | | | | |
| In a same-sex couple | -1.012*** | -0.345*** | -0.384*** | -1.108*** | -1.091*** | -1.194*** | -1.021*** |
|  | (0.123) | (0.122) | (0.122) | (0.123) | (0.125) | (0.123) | (0.123) |
| Observations | 5,210,836 | 5,210,836 | 5,210,836 | 5,210,836 | 4,571,778 | 5,210,836 | 5,210,836 |
| Mean of dependent variable | 27.675 | 27.675 | 27.675 | 27.675 | 27.796 | 27.675 | 27.675 |
| Adjusted $R^2$ | 0.020 | 0.045 | 0.043 | 0.021 | 0.022 | 0.022 | 0.020 |
| *Controls for:* | | | | | | | |
| State and year FE | ✓ | ✓ | ✓ | ✓ | ✓ | ✓ | ✓ |
| Demographic controls | ✓ | ✓ | ✓ | ✓ | ✓ | ✓ | ✓ |
| Partner/spouse controls | ✓ | ✓ | ✓ | ✓ | ✓ | ✓ | ✓ |
| Fertility and marital status | ✓ | ✓ | ✓ | ✓ | ✓ | ✓ | ✓ |
| Student and army status | ✓ | ✓ | ✓ | | | | |
| Occupation FE | | ✓ | | | | | |
| Industry FE | | | ✓ | | | | |
| Family income | | | | ✓ | | | |
| Urbanicity | | | | | ✓ | | |
| Homeownership | | | | | | ✓ | |
| LGBTQ+ policies | | | | | | | ✓ |

See also notes in Table 2. LGBTQ+ policies: constitutional and statutory bans on same-sex marriage, same-sex marriage legalization, same-sex domestic partnership legalization, same-sex civil union legalization, LGBTQ+ anti-discrimination laws, and LGBTQ+ hate crime laws Source: ACS 2008-2019. * $p < 0.10$, ** $p < 0.05$, *** $p < 0.01$.

**Table 8: Commuting time. By education level.**

|  | High-educ w/ high-educ partner | High-educ w/ low-educ partner | Low-educ w/ high-educ partner | Low-educ w/ low-educ partner |
|---|---|---|---|---|
|  | (1) | (2) | (3) | (4) |
| *Panel A: Women in SSC and DSC* | | | | |
| In a same-sex couple | 2.220*** | 0.598** | 2.128*** | 1.405*** |
|  | (0.193) | (0.278) | (0.292) | (0.183) |
| Observations | 1,155,948 | 670,662 | 425,499 | 2,159,300 |
| Mean of dependent variable | 24.088 | 25.091 | 21.984 | 22.403 |
| Adjusted $R^2$ | 0.042 | 0.025 | 0.026 | 0.018 |
| *Panel B: Men in SSC and DSC* | | | | |
| In a same-sex couple | -0.572*** | -0.538** | -1.313*** | -1.752*** |
|  | (0.212) | (0.267) | (0.290) | (0.220) |
| Observations | 1,385,856 | 571,011 | 671,106 | 2,582,863 |
| Mean of dependent variable | 27.684 | 27.771 | 27.670 | 27.652 |
| Adjusted $R^2$ | 0.043 | 0.028 | 0.020 | 0.014 |
| *Controls for:* | | | | |
| State and year FE | ✓ | ✓ | ✓ | ✓ |
| Demographic controls | ✓ | ✓ | ✓ | ✓ |
| Partner/spouse controls | ✓ | ✓ | ✓ | ✓ |
| Fertility and marital status | ✓ | ✓ | ✓ | ✓ |

See also notes in Table 2. Unlike Table 2, the indicators for education level are not included in "Demographic controls" and "Partner/spouse controls" since they are used to select the sub-samples across specifications. Source: ACS 2008-2019. * $p < 0.10$, ** $p < 0.05$, *** $p < 0.01$.

**Table 9: Within-couple commuting time gap. By sex and couple type.**

|  | (1) | (2) | (3) | (4) | (5) |
|---|---|---|---|---|---|
| *Panel A: Women in SSC and DSC* | | | | | |
| In a same-sex couple | -1.761*** | -2.028*** | -2.103*** | -2.102*** | -1.537*** |
|  | (0.143) | (0.142) | (0.142) | (0.142) | (0.145) |
| Observations | 3,613,685 | 3,613,685 | 3,613,685 | 3,613,685 | 3,613,685 |
| Mean of dependent variable | 17.157 | 17.157 | 17.157 | 17.157 | 17.157 |
| $R^2$ | 0.000 | 0.011 | 0.011 | 0.011 | 0.012 |
| | | | | | |
| *Panel B: Men in SSC and DSC* | | | | | |
| In a same-sex couple | -0.514*** | -1.000*** | -1.125*** | -1.078*** | -0.375** |
|  | (0.157) | (0.156) | (0.156) | (0.156) | (0.158) |
| Observations | 3,612,771 | 3,612,771 | 3,612,771 | 3,612,771 | 3,612,771 |
| Mean of dependent variable | 17.166 | 17.166 | 17.166 | 17.166 | 17.166 |
| $R^2$ | 0.000 | 0.011 | 0.011 | 0.011 | 0.012 |
| | | | | | |
| *Controls for:* | | | | | |
| State and year FE | | ✓ | ✓ | ✓ | ✓ |
| Demographic controls | | | ✓ | ✓ | ✓ |
| Partner/spouse controls | | | | ✓ | ✓ |
| Fertility and marital status | | | | | ✓ |

Only household heads have been included. Note: within-couple commute gap are the same for women and men in different-sex couples (by construction), so Panel A compares women in same-sex couples to both men and women in different-sex couples, while Panel B compares men in same-sex couples to both men and women in different-sex couples. See notes in Table 2. Source: ACS 2008-2019. * $p < 0.10$, ** $p < 0.05$, *** $p < 0.01$.

# Table 10: Hours worked. By sex and couple type.

|  | (1) | (2) | (3) | (4) | (5) | (6) |
|---|---|---|---|---|---|---|
| *Panel A: Women in SSC and DSC* | | | | | | |
| In a same-sex couple | 3.079*** | 2.941*** | 2.863*** | 2.144*** | 2.088*** | 2.448*** |
|  | (0.053) | (0.053) | (0.053) | (0.054) | (0.054) | (0.051) |
| Commuting time |  |  | 0.039*** |  | 0.038*** | 0.035*** |
|  |  |  | (0.000) |  | (0.000) | (0.000) |
| Observations | 4,411,409 | 4,411,409 | 4,411,409 | 4,411,409 | 4,411,409 | 4,411,409 |
| Mean of dependent variable | 37.766 | 37.766 | 37.766 | 37.766 | 37.766 | 37.766 |
| $R^2$ | 0.001 | 0.035 | 0.040 | 0.045 | 0.050 | 0.130 |
| *Panel B: Men in SSC and DSC* | | | | | | |
| In a same-sex couple | -1.831*** | -2.082*** | -2.050*** | -0.955*** | -0.936*** | -1.332*** |
|  | (0.056) | (0.055) | (0.055) | (0.057) | (0.057) | (0.053) |
| Commuting time |  |  | 0.017*** |  | 0.016*** | 0.018*** |
|  |  |  | (0.000) |  | (0.000) | (0.000) |
| Observations | 5,210,836 | 5,210,836 | 5,210,836 | 5,210,836 | 5,210,836 | 5,210,836 |
| Mean of dependent variable | 44.199 | 44.199 | 44.199 | 44.199 | 44.199 | 44.199 |
| $R^2$ | 0.000 | 0.019 | 0.020 | 0.023 | 0.024 | 0.112 |
| *Controls for:* | | | | | | |
| Hourly wages |  | ✓ | ✓ | ✓ | ✓ | ✓ |
| State and year FE |  | ✓ | ✓ | ✓ | ✓ | ✓ |
| Demographic controls |  | ✓ | ✓ | ✓ | ✓ | ✓ |
| Partner/spouse controls |  | ✓ | ✓ | ✓ | ✓ | ✓ |
| Fertility and marital status |  |  |  | ✓ | ✓ |  |
| Student and army status |  |  |  |  |  | ✓ |
| Occupation FE |  |  |  |  |  | ✓ |

The dependent variable is the number of hours per week that the individual usually worked (if the person worked during the 12 months preceding the interview). Individuals working part-time are included in the sample. Commuting time includes individuals working from home (commuting time imputed as zero). Only respondents with non-missing commuting time have been included in the analysis See also notes in Table 2. Source: ACS 2008-2019. * $p < 0.10$, ** $p < 0.05$, *** $p < 0.01$.

**Table 11: Hours worked. By marital status and fertility.**

|  | Married w/ children | Married w/o children | Unmarried w/ children | Unmarried w/o children |
|---|---|---|---|---|
|  | (1) | (2) | (3) | (4) |
| *Panel A: Women in SSC and DSC* | | | | |
| In a same-sex couple | 3.523*** | 1.812*** | 1.771*** | 0.993*** |
|  | (0.150) | (0.110) | (0.175) | (0.104) |
| Commuting time | 0.043*** | 0.032*** | 0.024*** | 0.020*** |
|  | (0.001) | (0.001) | (0.002) | (0.001) |
| Observations | 1,518,968 | 1,049,278 | 144,190 | 227,662 |
| Mean of dependent variable | 37.263 | 38.693 | 37.643 | 39.168 |
| $R^2$ | 0.048 | 0.045 | 0.050 | 0.052 |
| *Panel B: Men in SSC and DSC* | | | | |
| In a same-sex couple | -2.256*** | -1.494*** | -1.343*** | -1.029*** |
|  | (0.203) | (0.115) | (0.326) | (0.094) |
| Commuting time | 0.016*** | 0.018*** | 0.022*** | 0.024*** |
|  | (0.000) | (0.001) | (0.002) | (0.001) |
| Observations | 1,972,381 | 1,092,622 | 166,510 | 235,897 |
| Mean of dependent variable | 44.678 | 43.949 | 42.682 | 42.748 |
| $R^2$ | 0.024 | 0.014 | 0.024 | 0.020 |
| *Controls for:* | | | | |
| Hourly wages | ✓ | ✓ | ✓ | ✓ |
| State and year FE | ✓ | ✓ | ✓ | ✓ |
| Demographic controls | ✓ | ✓ | ✓ | ✓ |
| Partner/spouse controls | ✓ | ✓ | ✓ | ✓ |

See also notes in Table 10. Source: ACS 2012-2019. * $p < 0.10$, ** $p < 0.05$, *** $p < 0.01$.

**Table 12: Log of hourly wages. By sex and couple type.**

|  | (1) | (2) | (3) | (4) | (5) | (6) | (7) | (8) |
|---|---|---|---|---|---|---|---|---|
| *Panel A: Women in SSC and DSC* | | | | | | | | |
| In a same-sex couple | 0.074*** | 0.028*** | 0.025*** | 0.076*** | 0.073*** | 0.023*** | 0.030*** | 0.028*** |
|  | (0.004) | (0.003) | (0.003) | (0.003) | (0.003) | (0.003) | (0.003) | (0.003) |
| Commuting time |  |  | 0.002*** |  | 0.002*** | 0.002*** |  | 0.002*** |
|  |  |  | (0.000) |  | (0.000) | (0.000) |  | (0.000) |
| Observations | 2,797,527 | 2,797,527 | 2,797,527 | 2,797,527 | 2,797,527 | 2,797,527 | 4,145,921 | 4,145,921 |
| Mean of dependent variable | 2.748 | 2.748 | 2.748 | 2.748 | 2.748 | 2.748 | 2.676 | 2.676 |
| $R^2$ | 0.000 | 0.252 | 0.260 | 0.256 | 0.264 | 0.413 | 0.390 | 0.394 |
| | | | | | | | | |
| *Panel B: Men in SSC and DSC* | | | | | | | | |
| In a same-sex couple | 0.049*** | -0.060*** | -0.057*** | 0.047*** | 0.049*** | -0.047*** | -0.050*** | -0.048*** |
|  | (0.004) | (0.003) | (0.003) | (0.003) | (0.003) | (0.003) | (0.003) | (0.003) |
| Commuting time |  |  | 0.002*** |  | 0.002*** | 0.001*** |  | 0.001*** |
|  |  |  | (0.000) |  | (0.000) | (0.000) |  | (0.000) |
| Observations | 4,356,263 | 4,356,263 | 4,356,263 | 4,356,263 | 4,356,263 | 4,356,263 | 4,813,049 | 4,813,049 |
| Mean of dependent variable | 2.955 | 2.955 | 2.955 | 2.955 | 2.955 | 2.955 | 2.925 | 2.925 |
| $R^2$ | 0.000 | 0.260 | 0.264 | 0.269 | 0.273 | 0.388 | 0.378 | 0.380 |
| | | | | | | | | |
| *Controls for:* | | | | | | | | |
| State and year FE |  | ✓ | ✓ | ✓ | ✓ | ✓ | ✓ | ✓ |
| Demographic controls |  | ✓ | ✓ | ✓ | ✓ | ✓ | ✓ | ✓ |
| Partner/spouse controls |  | ✓ | ✓ | ✓ | ✓ | ✓ | ✓ | ✓ |
| Fertility and marital status |  |  |  | ✓ | ✓ |  |  |  |
| Student and army status |  |  |  |  |  | ✓ | ✓ | ✓ |
| Occupation FE |  |  |  |  |  | ✓ | ✓ | ✓ |
| | | | | | | | | |
| Include part-time workers |  |  |  |  |  |  | ✓ | ✓ |

The dependent variable is the logarithm of the respondent's total pre-tax wage and salary income in the 12 months preceding the ACS interview divided by the estimated number of hours worked in the same 12 months. All wages have been adjusted for inflation using the FRED Consumer Price Index for All Urban Consumers (All Items). Respondents whose hourly wage was above the 99th percentile of the hourly wage distribution for the relevant sample have been excluded. Only respondents with a positive hourly wage and working at least 40h/week have been included in the analysis. Both married and unmarried couples included in this sample. Commuting time includes individuals working from home (commuting time imputed as zero). Only respondents with non-missing commuting time have been included in the analysis See also notes in Table 2. Source: ACS 2008-2019. * $p < 0.10$, ** $p < 0.05$, *** $p < 0.01$.

**Online Appendix for "Commuting to work and gender-conforming social norms: evidence from same-sex couples" (NOT MEANT FOR PUBLICATION)**

**Appendix A. Variable description**

**A.1 ACS Variables**

**A.1.1 Dependent variables**

*Commuting time* reports the total amount of time, in minutes, that it usually took the respondent to get from home to work in the week preceding the ACS interview. This variable is set to missing for individuals who did not work in such week. Unless otherwise specified, commuting time includes individuals working from home (commuting time imputed as zero).

*Within-couple commuting time gap* reports the (absolute value of the) difference in commuting times between the household head and their spouse or unmarried partner. Commuting time includes individuals working from home (commuting time imputed as zero). Couples in which at least one member was not working at the time of the ACS interview have been coded as missing.

*Total couple commuting time* reports total commuting time of the household head and their spouse or unmarried partner. Commuting time includes individuals working from home (commuting time imputed as zero). Couples in which at least one member was not working at the time of the ACS interview have been coded as missing.

*Number of hours worked weekly.* The ACS reports the number of hours per week that the respondent usually worked, if the person worked during the 12 months preceding the interview. This variable is top coded at 99. Respondents who did not work in the 12 months preceding the interview are assigned value zero.

*Wage and salary income* reports individual hourly pre-tax wage and salary income. Individuals were asked the usual number of hours worked in a week in the 12 months preceding the interview, the number of weeks worked in the 12 months preceding the interview (including paid vacation, paid sick leave, and military service), and the total pre-tax wage and salary income - that is, money received as an employee - for the 12 months preceding the interview. Given this definition of income, self-employed individuals have been excluded from this analysis. These three variables have been used to compute hourly earnings for each respondent. Since the number of weeks worked in the previous years is recorded as a categorical variable, it has been assumed that the actual number of weeks worked is the median of the selected interval. For instance, if the individual reported working 27-39 weeks, it has been assumed that she worked 33 weeks. Whenever indicated, we have adjusted income for inflation using the average annual FRED Consumer Price Index for All Urban Consumers (All Items).[2]

**A.1.2 Key independent variable: In a same-sex couple**

The ACS does not directly ask individuals about their sexual orientation. However, the ACS identifies a primary reference person, defined as "the person living or staying here in whose name this house or apartment is owned, being bought, or rented". The ACS also collects information on the relationship to the primary reference person for all members of the household, and the range of possible relationships includes husband, wife, and unmarried partner (as a different category than roommate or other nonrelative). By combining such information, it has been possible to create an indicator variable equal to one if an individual was in a same-sex couple; zero if an individual was in a different-sex couple. Both individuals married to a same-sex spouse and individuals living with a same-sex unmarried partner have been coded as individuals in same-sex couples.

It is worth nothing that, in order to reduce measurement error, in 2019 the ACS survey question explicitly distinguished between "opposite-sex husband/wife/spouse", "opposite-sex unmarried partner", "same-sex husband/wife/spouse", and "same-sex unmarried partner". In addition, the options for unmarried partners were moved higher in the list of potential relation categories, thus increasing its salience.

**A.1.3 Additional individual variables**

*Sex* reports whether the person was male or female. Note that sex in the ACS is reported as a binary variable.

*Age* reports a person's age in years at the time of the interview. A similar variable has been constructed to report the age of a person's spouse or unmarried partner.

*Race* includes a series of indicator variables constructed to record a person's race: White, Black, Asian, or other races. The indicator *Asian* includes Chinese, Japanese, Other Asian or Pacific Islander. The indicator *other races* includes American Indian, Alaska Native, other race not listed, or individuals who selected two or three major races. A similar set of variables has been constructed to report the race of a person's spouse or unmarried partner.

---

[2] Source: https://fred.stlouisfed.org/series/CPIAUCSL.

*Hispanic* is an indicator equal to one if a person self-identified as Mexican, Puerto Rican, Cuban, or Other Hispanic; zero otherwise. A similar variable has been constructed to report the ethnicity of a person's spouse or unmarried partner.

*Higher Education* is an indicator equal to one if a person's highest degree completed was a Bachelor's degree or higher (Master's degree, Professional degree beyond a Bachelor's degree, Doctoral degree); zero otherwise. A similar variable has been constructed to report the education level of a person's spouse or unmarried partner.

*Number of children* reports the number of own children (of any age or marital status) residing with each individual. This variable includes step-children and adopted children as well as biological children. This variable is coded as zero for people with no children present in the household.

*Number of children under age 5* reports the number of own children age 4 or under residing with each individual. This variable includes step-children and adopted children as well as biological children. This variable is coded as zero for people with no children under 5 present in the household.

*Married* is an indicator equal to one if a person is a member of a (same-sex or different-sex) married couple; zero otherwise. Married same-sex couples can be identified from 2012. Before 2012, married same-sex couples were re-coded as unmarried same-sex couples by the ACS.

*Student status* is an indicator equal to one if a person attended school or college in the 3 months preceding the interview; zero otherwise.

*In the army* is an indicator equal to one if a person reported being employed in the Armed forces (including "Armed forces: at work" and "Armed forces: with job but not at work"); zero otherwise.

*Employed* is an indicator equal to one if a person was working in the week preceding the interview; zero otherwise.

*In the labor force* is an indicator equal to one if a person was a part of the labor force, either working or seeking work, in the week preceding the interview; zero if a person was out of the labor force, or did not have a job, was looking for a job, but had not yet found one at the time of the interview.

*Total family income* reports the total pre-tax money income earned by one's family from all sources for the 12 months preceding the interview. Amounts are expressed in contemporary dollars, and not adjusted for inflation.

*Occupation* records a person's primary occupation using the IPUMS harmonized occupation coding based on the Census Bureau's 2010 ACS occupation classification scheme. Unemployed persons were to give their most recent occupation, if they had worked in the 5 years preceding the interview, otherwise they were classified as "Unemployed, with No Work Experience in the Last 5 Years or Earlier or Never Worked".

*Industry* reports the type of industry in which the person performed an occupation using the IPUMS harmonized industry coding based on the 1990 Census Bureau industrial classification scheme. Unemployed persons were to give their most recent industry, if they had worked in the 5 years preceding the interview, otherwise they were classified as "N/A (not applicable)" or "Last worked 1984 or earlier".

*Urbanicity* reports whether a person resided in a metropolitan area (including both inside and outside the central/principal city). Persons with indeterminable metropolitan status were recorded as missing. Indeed, confidentiality requirements have limited the details regarding metropolitan status for some individuals.

**A.1.4 LGBTQ+ policy variables.**

*SSM legal* is an indicator variable equal to one in all states and time periods when same-sex marriage was legal; zero otherwise. The effective date has been used to code this variable. These data have been primarily obtained from the National Center for Lesbian Rights.[3]

*SSM ban* is a series of indicator variables equal to one in all states and time periods in which same-sex marriage was banned in the state constitution or state statute; zero otherwise. These indicators remain equal to one even in later years after the legalization of same-sex marriage in a given state. When more than one statutory ban was passed in a state, the oldest one has been used to code the

---

[3] Source: http://www.nclrights.org/wp-content/uploads/2015/07/Relationship-Recognition.pdf. Accessed Oct/1/2019.

state statute ban variable. These data have been primarily obtained from the Freedom to Marry campaign.[4]

*Domestic partnership* is an indicator variable equal to one in all states and time periods in which same-sex domestic partnerships were legal; zero otherwise. This indicator remains equal to one even in later years when\if a state had converted same-sex domestic partnerships into marriages. These data have been primarily obtained from the National Center for Lesbian Rights.[5]

*Civil union* is an indicator variable equal to one in all states and time periods in which same-sex civil unions were legal; zero otherwise. This indicator remains equal to one even in later years when\if a state had converted same-sex civil unions in marriages. These data have been primarily obtained from the National Center for Lesbian Rights.[6]

*Anti-discrimination law* is an indicator equal to one in all states and time periods in which employer discrimination based on sexual orientation was not allowed; zero otherwise. This variable has been set equal to one even if the law covered only sexual orientation, not gender identity, or if a law protecting trans individuals was passed at a later date. Laws protecting only public employees have not been considered. These data have been primarily obtained from the Freedom for All Americans campaign.[7]

*Hate crime* is a series of indicator variables equal to one in all states and time periods in which there was a law specifically addressing hate or bias crimes based on sexual orientation only, or on sexual orientation and gender identity; zero otherwise. Since some states passed these laws after 2009, these variables have not been set equal to one for all states after President Obama signed the Matthew Shepard and James Byrd, Jr. Hate Crimes Prevention Act into law on October 28, 2009. These data have been primarily obtained from the Human Rights Campaign.[8]

---

[4] Source: http://www.freedomtomarry.org/pages/winning-in-the-states. Accessed Oct/1/2019.
[5] Source: http://www.nclrights.org/wp-content/uploads/2015/07/Relationship-Recognition.pdf. Accessed Oct/1/2019.
[6] Source: http://www.nclrights.org/wp-content/uploads/2015/07/Relationship-Recognition.pdf. Accessed Oct/1/2019.
[7] Source: https://www.freedomforallamericans.org/states/.Accessed: Oct/21/2019. We have subsequently checked that no other law was passed in 2019.
[8] Source: https://www.hrc.org/state-maps/hate-crimes. Accessed: Oct/25/2019. We have subsequently checked that no other law was passed in 2019

# Appendix B. ACS additional tables and figures

**Figure B1: Raw gaps in commuting time (including working from home) for individuals in same-sex vs. individuals in different-sex couples.**

**Panel A: Women.**

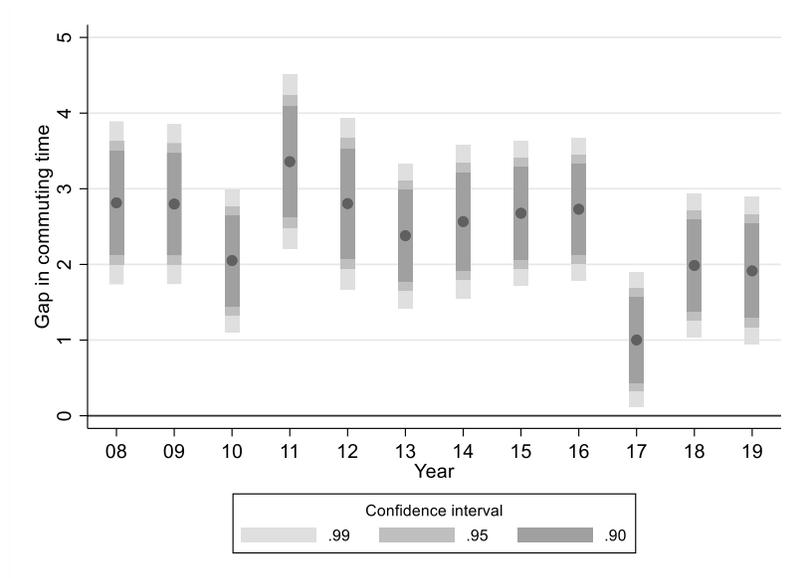

**Panel B: Men.**

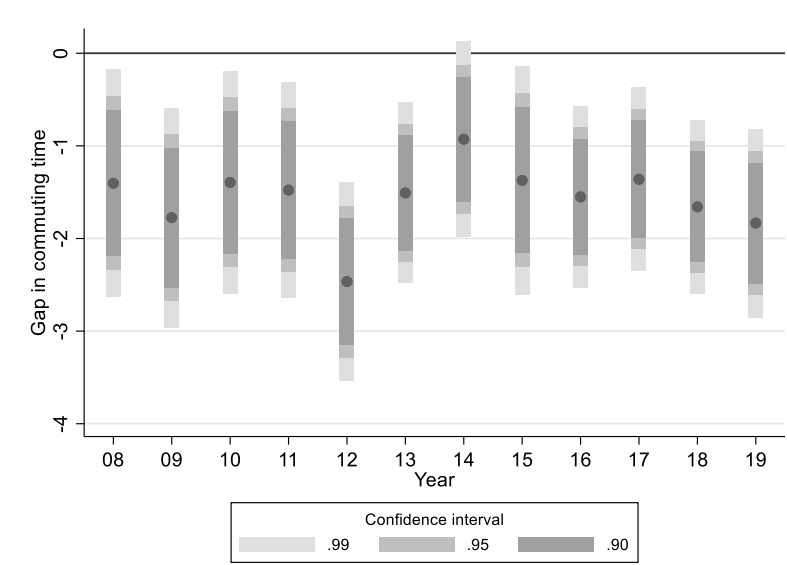

These figures report the estimated gap between women (men in Panel B) in same-sex couples and women (men) in different-sex couples from 12 different regressions, one for each year. The dependent variable is commuting time (including individuals working from home). See also notes in Table 1. As in Table 1 Column 1, these regressions do not include demographic controls, partner/spouse controls, marital status, fertility, or state fixed effects. Source: ACS 2008-2019.

**Figure B2: Within-couple commuting time gap distribution.**

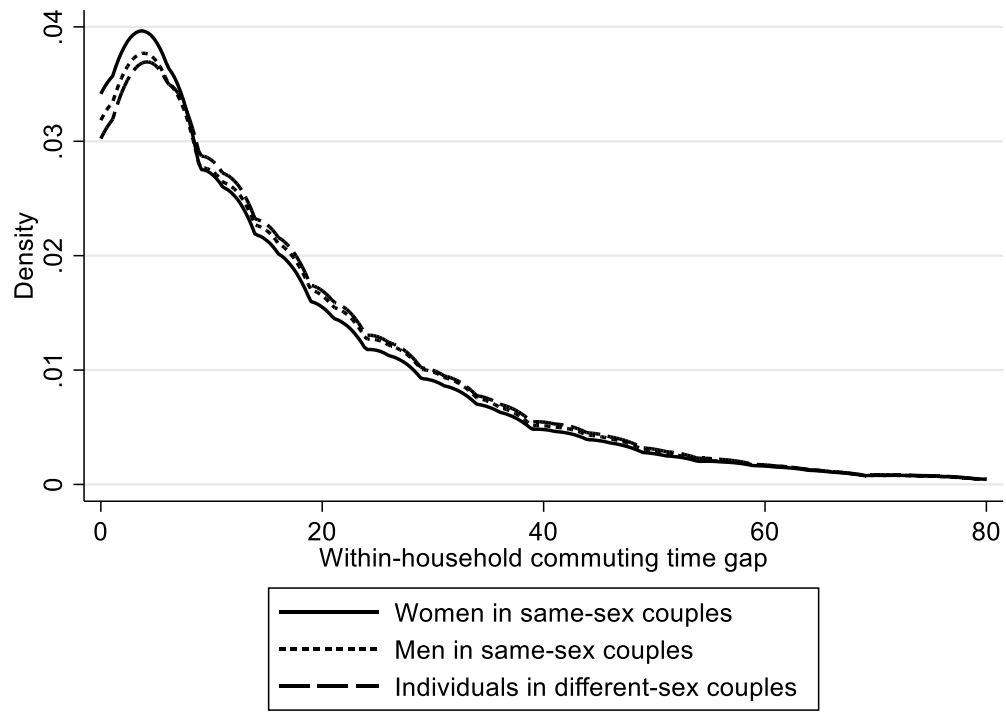

Respondents younger than 18 or older than 64 have been excluded. Commuting time gap censored at 80 minutes. Bin width equal to 8 minutes. Unweighted statistics. Source: ACS 2008-2019.

**Table B1: ACS sample sizes. Individuals 18-64 in same-sex and different-sex couples.**

|      | Same-sex | | Different-sex | |
|------|--------|--------|-----------|-----------|
|      | Female | Male   | Married   | Unmarried |
| 2008 | 5,453  | 5,079  | 997,747   | 96,396    |
| 2009 | 5,703  | 5,285  | 994,337   | 99,090    |
| 2010 | 5,733  | 5,340  | 977,773   | 106,248   |
| 2011 | 5,834  | 5,384  | 945,122   | 104,172   |
| 2012 | 6,080  | 5,603  | 942,970   | 106,056   |
| 2013 | 6,982  | 6,791  | 944,980   | 111,931   |
| 2014 | 7,380  | 7,110  | 929,088   | 113,035   |
| 2015 | 8,061  | 7,723  | 927,944   | 116,554   |
| 2016 | 8,036  | 8,021  | 922,524   | 116,246   |
| 2017 | 8,871  | 8,314  | 926,510   | 121,186   |
| 2018 | 9,137  | 8,975  | 922,169   | 122,709   |
| 2019 | 9,167  | 8,737  | 922,234   | 126,020   |
| Total| 86,437 | 82,362 | 11,353,398| 1,339,643 |

Notes: Sample includes all respondents (both primary reference person and unmarried partner or married spouse) in a same-sex or different-sex married/unmarried couple. Respondents younger than 18 or older than 64 have been excluded. Source: ACS 2008-2019.

**Table B2: ACS sample sizes. Individuals 18-64 in married/unmarried couples.**

|      | Female same-sex | | Male same-sex | | Different-sex | |
|------|---------|-----------|---------|-----------|---------|-----------|
|      | Married | Unmarried | Married | Unmarried | Married | Unmarried |
| 2012 | 1,682   | 4,398     | 1,412   | 4,191     | 942,970 | 106,056   |
| 2013 | 2,383   | 4,599     | 2,095   | 4,696     | 944,980 | 111,931   |
| 2014 | 2,909   | 4,471     | 2,891   | 4,219     | 929,088 | 113,035   |
| 2015 | 4,047   | 4,014     | 3,588   | 4,135     | 927,944 | 116,554   |
| 2016 | 4,374   | 3,662     | 4,182   | 3,839     | 922,524 | 116,246   |
| 2017 | 5,296   | 3,575     | 4,681   | 3,633     | 926,510 | 121,186   |
| 2018 | 5,429   | 3,708     | 5,140   | 3,835     | 922,169 | 122,709   |
| 2019 | 5,453   | 3,714     | 4,958   | 3,779     | 922,234 | 126,020   |

Notes: Sample includes all respondents (both primary reference person and unmarried partner or married spouse) in a same-sex or different-sex married/unmarried couple. Respondents younger than 18 or older than 64 have been excluded. Marital status recorded in the ACS for same-sex couples only from 2012. Source: ACS 2012-2019.

**Table B3: Descriptive statistics – additional variable.**

|  | Women | | Men | |
|---|---|---|---|---|
|  | Same-sex couples | Different-sex couples | Same-sex couples | Different-sex couples |
| Variable | (1) | (2) | (3) | (4) |
| Age | 42.303 | 44.069 | 43.542 | 45.096 |
| White | 0.806 | 0.794 | 0.818 | 0.791 |
| Black | 0.095 | 0.074 | 0.064 | 0.085 |
| Asian | 0.027 | 0.066 | 0.045 | 0.055 |
| Other race | 0.072 | 0.067 | 0.073 | 0.069 |
| Hispanic | 0.126 | 0.146 | 0.148 | 0.152 |
| Bachelor's degree | 0.447 | 0.361 | 0.488 | 0.340 |
| Has child | 0.326 | 0.599 | 0.137 | 0.621 |
| Has child age 0-4 | 0.088 | 0.188 | 0.043 | 0.201 |
| Married | 0.483 | 0.876 | 0.460 | 0.870 |
| Student | 0.095 | 0.061 | 0.076 | 0.041 |
| In the army | 0.004 | 0.001 | 0.003 | 0.009 |
| Employed | 0.810 | 0.680 | 0.821 | 0.858 |
| In the labor force | 0.849 | 0.715 | 0.860 | 0.895 |
| Total family income | 74,282 | 101,340 | 101,851 | 101,117 |
| Weekly hours worked | 40.758 | 37.644 | 42.335 | 44.188 |
| N | 86,437 | 6,554,055 | 82,362 | 6,138,986 |

Weighted means. Sample size (N) refers to the total number of respondents in the relevant sub-group (i.e., individuals in same-sex or different-sex couples). Weekly hours worked reported only for those working in the week preceding the ACS interview. Respondents younger than 18 or older than 64 have been excluded. Source: ACS 2008-2019 (2012-2019 for marital status). All differences are statistically significant at the 1-percent level.

**Table B4: Commuting time (including working from home). By education level, with controls for occupation.**

|  | High-educ w/ high-educ partner | High-educ w/ low-educ partner | Low-educ w/ high-educ partner | Low-educ w/ low-educ partner |
|---|---|---|---|---|
|  | (1) | (2) | (3) | (4) |
| *Panel A: Women in SSC and DSC* | | | | |
| In a same-sex couple | 1.894*** | 0.204 | 1.672*** | 0.974*** |
|  | (0.193) | (0.276) | (0.293) | (0.183) |
| Observations | 1,155,948 | 670,662 | 425,499 | 2,159,300 |
| Mean of dependent variable | 24.088 | 25.091 | 21.984 | 22.403 |
| Adjusted $R^2$ | 0.066 | 0.049 | 0.058 | 0.041 |
| | | | | |
| *Panel B: Men in SSC and DSC* | | | | |
| In a same-sex couple | -0.298 | -0.299 | -0.326 | -0.489** |
|  | (0.212) | (0.267) | (0.291) | (0.219) |
| Observations | 1,385,856 | 571,011 | 671,106 | 2,582,863 |
| Mean of dependent variable | 27.684 | 27.771 | 27.670 | 27.652 |
| Adjusted $R^2$ | 0.064 | 0.053 | 0.044 | 0.042 |
| | | | | |
| *Controls for:* | | | | |
| State and year FE | ✓ | ✓ | ✓ | ✓ |
| Demographic controls | ✓ | ✓ | ✓ | ✓ |
| Partner/spouse controls | ✓ | ✓ | ✓ | ✓ |
| Fertility and marital status | ✓ | ✓ | ✓ | ✓ |
| Occupation FE | ✓ | ✓ | ✓ | ✓ |

See also notes in Table 8. Source: ACS 2008-2019. * $p < 0.10$, ** $p < 0.05$, *** $p < 0.01$.

# Table B5: Commuting time (including working from home). Additional restrictions.

|  | No students | No army | 2012-2019 | Robust SE | No weights |
|---|---|---|---|---|---|
|  | (1) | (2) | (3) | (4) | (5) |
| *Panel A: Women in SSC and DSC* | | | | | |
| In a same-sex couple | 1.770*** | 1.762*** | 1.641*** | 1.761*** | 1.829*** |
|  | (0.119) | (0.114) | (0.134) | (0.103) | (0.099) |
| Observations | 4,139,712 | 4,404,566 | 2,940,098 | 4,411,409 | 4,411,409 |
| Mean of dependent variable | 23.166 | 23.197 | 23.552 | 23.201 | 23.201 |
| Adjusted $R^2$ | 0.027 | 0.027 | 0.027 | 0.027 | 0.023 |
|  | | | | | |
| *Panel B: Men in SSC and DSC* | | | | | |
| In a same-sex couple | -1.017*** | -1.010*** | -1.263*** | -1.021*** | -1.156*** |
|  | (0.128) | (0.123) | (0.143) | (0.113) | (0.105) |
| Observations | 5,007,438 | 5,154,713 | 3,467,410 | 5,210,836 | 5,210,836 |
| Mean of dependent variable | 27.721 | 27.703 | 27.976 | 27.675 | 27.675 |
| Adjusted $R^2$ | 0.020 | 0.020 | 0.021 | 0.020 | 0.018 |
|  | | | | | |
| *Controls for:* | | | | | |
| State and year FE | ✓ | ✓ | ✓ | ✓ | ✓ |
| Demographic controls | ✓ | ✓ | ✓ | ✓ | ✓ |
| Partner/spouse controls | ✓ | ✓ | ✓ | ✓ | ✓ |
| Fertility and marital status | ✓ | ✓ | ✓ | ✓ | ✓ |

See also notes in Table 2. Source: ACS 2008-2019 (2012-2019 in Column 3). * $p < 0.10$, ** $p < 0.05$, *** $p < 0.01$.

**Table B6: Commuting time (including working from home). By sex and couple type. Only working individuals with working partner/spouse.**

|  | (1) | (2) | (3) | (4) | (5) |
|---|---|---|---|---|---|
| *Panel A: Women in SSC and DSC* | | | | | |
| In a same-sex couple | 2.419*** | 2.054*** | 1.983*** | 2.045*** | 1.600*** |
|  | (0.124) | (0.122) | (0.122) | (0.122) | (0.125) |
| Observations | 3,702,772 | 3,702,772 | 3,702,772 | 3,702,772 | 3,702,772 |
| Mean of dependent variable | 23.051 | 23.051 | 23.051 | 23.051 | 23.051 |
| $R^2$ | 0.000 | 0.020 | 0.026 | 0.027 | 0.028 |
| | | | | | |
| *Panel B: Men in SSC and DSC* | | | | | |
| In a same-sex couple | -1.102*** | -1.858*** | -1.734*** | -1.819*** | -0.886*** |
|  | (0.134) | (0.132) | (0.132) | (0.132) | (0.136) |
| Observations | 3,613,617 | 3,613,617 | 3,613,617 | 3,613,617 | 3,613,617 |
| Mean of dependent variable | 27.218 | 27.218 | 27.218 | 27.218 | 27.218 |
| $R^2$ | 0.000 | 0.020 | 0.020 | 0.020 | 0.021 |
| | | | | | |
| *Controls for:* | | | | | |
| State and year FE | | ✓ | ✓ | ✓ | ✓ |
| Demographic controls | | | ✓ | ✓ | ✓ |
| Partner/spouse controls | | | | ✓ | ✓ |
| Fertility and marital status | | | | | ✓ |

See also notes in Table 2. Source: ACS 2008-2019. * $p < 0.10$, ** $p < 0.05$, *** $p < 0.01$.

**Table B7: Commuting time (including working from home). Sub-sample analysis. Only working individuals with working partner/spouse.**

|  | Married w/ children | Married w/o children | Unmarried w/ children | Unmarried w/o children |
|---|---|---|---|---|
|  | (1) | (2) | (3) | (4) |
| *Panel A: Women in SSC and DSC* | | | | |
| In a same-sex couple | 2.398*** | 1.729*** | 0.882** | 0.339 |
|  | (0.352) | (0.273) | (0.409) | (0.232) |
| Observations | 1,344,196 | 825,535 | 118,422 | 192,777 |
| Mean of dependent variable | 23.220 | 23.447 | 23.781 | 23.308 |
| $R^2$ | 0.029 | 0.028 | 0.026 | 0.035 |
|  | | | | |
| *Panel B: Men in SSC and DSC* | | | | |
| In a same-sex couple | -1.627*** | -1.077*** | -1.902** | -0.899*** |
|  | (0.544) | (0.277) | (0.847) | (0.216) |
| Observations | 1,324,583 | 783,957 | 113,986 | 195,681 |
| Mean of dependent variable | 28.221 | 26.829 | 27.197 | 25.815 |
| $R^2$ | 0.023 | 0.020 | 0.016 | 0.021 |
|  | | | | |
| *Controls for:* | | | | |
| State and year FE | ✓ | ✓ | ✓ | ✓ |
| Demographic controls | ✓ | ✓ | ✓ | ✓ |
| Partner/spouse controls | ✓ | ✓ | ✓ | ✓ |

See also notes in Table 2. Source: ACS 2012-2019. * $p < 0.10$, ** $p < 0.05$, *** $p < 0.01$.

**Table B8: Commuting time (excluding working from home). By sex and couple type.**

|  | (1) | (2) | (3) | (4) | (5) |
|---|---|---|---|---|---|
| *Panel A: Women in SSC and DSC* | | | | | |
| In a same-sex couple | 2.491*** | 2.048*** | 1.968*** | 2.025*** | 1.873*** |
|  | (0.116) | (0.113) | (0.113) | (0.113) | (0.115) |
| Observations | 4,151,189 | 4,151,189 | 4,151,189 | 4,151,189 | 4,151,189 |
| Mean of dependent variable | 24.619 | 24.619 | 24.619 | 24.619 | 24.619 |
| $R^2$ | 0.000 | 0.023 | 0.030 | 0.030 | 0.031 |
| | | | | | |
| *Panel B: Men in SSC and DSC* | | | | | |
| In a same-sex couple | -0.874*** | -1.665*** | -1.632*** | -1.740*** | -0.633*** |
|  | (0.125) | (0.123) | (0.123) | (0.123) | (0.126) |
| Observations | 4,967,002 | 4,967,002 | 4,967,002 | 4,967,002 | 4,967,002 |
| Mean of dependent variable | 29.032 | 29.032 | 29.032 | 29.032 | 29.032 |
| $R^2$ | 0.000 | 0.021 | 0.022 | 0.022 | 0.023 |
| | | | | | |
| *Controls for:* | | | | | |
| State and year FE | | ✓ | ✓ | ✓ | ✓ |
| Demographic controls | | | ✓ | ✓ | ✓ |
| Partner/spouse controls | | | | ✓ | ✓ |
| Fertility and marital status | | | | | ✓ |

See notes in Table 2. $^*p<0.10$, $^{**}p<0.05$, $^{***}p<0.01$.

**Table B9: Within-couple commuting time gap. Sub-sample analysis.**

|  | Married w/ children | Married w/o children | Unmarried w/ children | Unmarried w/o children |
|---|---|---|---|---|
|  | (1) | (2) | (3) | (4) |
| *Panel A: Women in SSC and DSC* | | | | |
| In a same-sex couple | -1.221*** | -0.860*** | -3.163*** | -2.349*** |
|  | (0.407) | (0.331) | (0.462) | (0.257) |
| Observations | 1,330,012 | 789,268 | 115,552 | 183,147 |
| Mean of dependent variable | 17.965 | 16.938 | 17.051 | 16.132 |
| $R^2$ | 0.013 | 0.012 | 0.010 | 0.014 |
| | | | | |
| *Panel B: Men in SSC and DSC* | | | | |
| In a same-sex couple | -0.121 | -0.815*** | -1.534 | -0.670*** |
|  | (0.662) | (0.287) | (0.967) | (0.256) |
| Observations | 1,328,182 | 790,295 | 113,574 | 185,504 |
| Mean of dependent variable | 17.967 | 16.941 | 17.100 | 16.215 |
| $R^2$ | 0.013 | 0.012 | 0.010 | 0.013 |
| | | | | |
| *Controls for:* | | | | |
| State and year FE | ✓ | ✓ | ✓ | ✓ |
| Demographic controls | ✓ | ✓ | ✓ | ✓ |
| Partner/spouse controls | ✓ | ✓ | ✓ | ✓ |

See also notes in Table 9. Source: ACS 2012-2019. * $p < 0.10$, ** $p < 0.05$, *** $p < 0.01$.

**Table B10: Hours worked. By marital status and fertility. No control for commuting time.**

|  | Married w/ children | Married w/o children | Unmarried w/ children | Unmarried w/o children |
|---|---|---|---|---|
|  | (1) | (2) | (3) | (4) |
| *Panel A: Women in SSC and DSC* | | | | |
| In a same-sex couple | 3.627*** | 1.862*** | 1.795*** | 1.002*** |
|  | (0.150) | (0.110) | (0.175) | (0.105) |
| Observations | 1,518,968 | 1,049,278 | 144,190 | 227,662 |
| Mean of dependent variable | 37.263 | 38.693 | 37.643 | 39.168 |
| $R^2$ | 0.042 | 0.041 | 0.047 | 0.050 |
| *Panel B: Men in SSC and DSC* | | | | |
| In a same-sex couple | -2.281*** | -1.516*** | -1.390*** | -1.058*** |
|  | (0.204) | (0.115) | (0.326) | (0.094) |
| Observations | 1,972,381 | 1,092,622 | 166,510 | 235,897 |
| Mean of dependent variable | 44.678 | 43.949 | 42.682 | 42.748 |
| $R^2$ | 0.023 | 0.012 | 0.021 | 0.017 |
| *Controls for:* | | | | |
| Hourly wages | ✓ | ✓ | ✓ | ✓ |
| State and year FE | ✓ | ✓ | ✓ | ✓ |
| Demographic controls | ✓ | ✓ | ✓ | ✓ |
| Partner/spouse controls | ✓ | ✓ | ✓ | ✓ |

See also notes in Table 10. Source: ACS 2012-2019. * $p < 0.10$, ** $p < 0.05$, *** $p < 0.01$.